\documentclass{aa}  

\usepackage{graphicx}
\usepackage{txfonts}
\usepackage{microtype}
\usepackage{booktabs}
\usepackage{multirow}
\usepackage{placeins}

\usepackage[colorlinks=true,allcolors=blue]{hyperref}

\usepackage{amsmath}
\usepackage{amssymb}
\usepackage{mathtools}
\DeclareMathOperator*{\argmax}{arg\,max\,}
\DeclareMathOperator*{\argmin}{arg\,min\,}
\newcommand{\vsini}{$v_\mathrm{e}\sin i$}
\newcommand{\kms}{km\,s$^{-1}$}
\newcommand{\sco}{$\tau$~Sco}

\newcommand{\orcidlink}[1]{\protect\href{https://orcid.org/#1}{\protect\includegraphics[width=8pt]{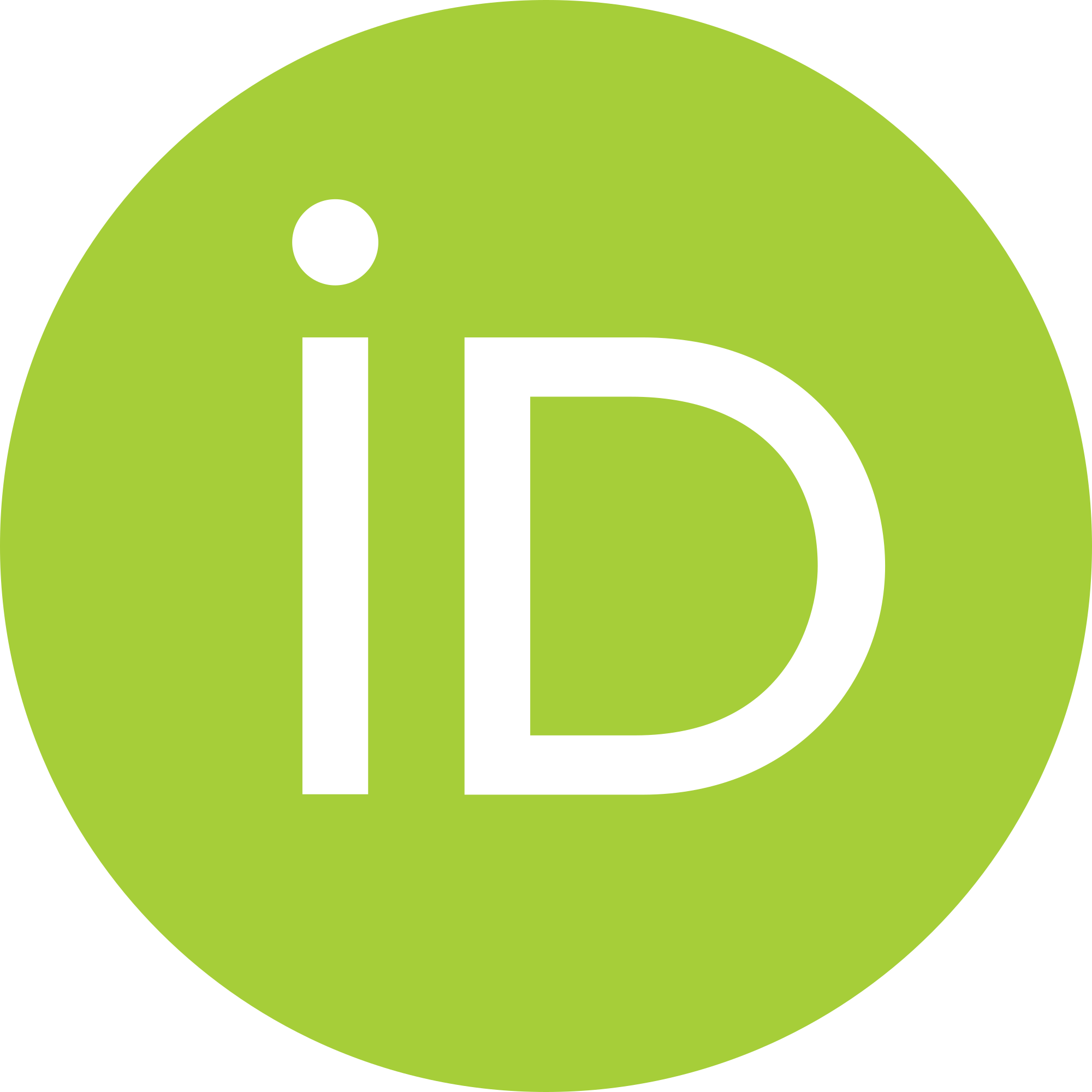}}}
\newcommand{\cla}[1]{\textcolor{black}{#1}}
\newcommand{\revb}[1]{\textcolor{black}{#1}}
\usepackage{placeins}
\begin{document}
\makeatletter
\let\linenumbers\relax
\let\endlinenumbers\relax
\makeatother
\pdfimageresolution=300
\title{Probabilistic Zeeman-Doppler imaging of stellar magnetic fields}

\subtitle{I. Analysis of $\tau$~Scorpii in the weak-field limit}

\author{
J.R. Andersson\inst{1}\orcidlink{0000-0002-6811-2776}
\and
O. Kochukhov\inst{2}\orcidlink{0000-0003-3061-4591}
\and
Z. Zhao\inst{3}\orcidlink{0000-0002-0368-786X}
\and
J. Sj\"olund\inst{1}\orcidlink{0000-0002-9099-3522}
}

\authorrunning{Andersson, J.R., et al.} 

\institute{
Department of Information Technology, Uppsala University, Box 337, SE-75120 Uppsala, Sweden\\ \email{jennifer.andersson@it.uu.se}
\and
Department of Physics and Astronomy, Uppsala University, Box 516, SE-75120 Uppsala, Sweden
\and
Department of Computer and Information Science, Link\"{o}ping University, SE-58183 Link\"{o}ping, Sweden
}

\date{Received ; accepted }

\abstract
{Zeeman-Doppler imaging (ZDI) is used to study the surface magnetic field topology of stars, based on high-resolution spectropolarimetric time series observations. Multiple ZDI inversions have been conducted for the early B-type star \sco, which has been found to exhibit a weak but complex non-dipolar surface magnetic field.}
{The classical ZDI framework suffers from a significant limitation in that it provides little to no reliable uncertainty quantification for the reconstructed magnetic field maps, with essentially all published results being confined to point estimates. To fill this gap, we propose a Bayesian framework for probabilistic ZDI. Here, the proposed framework is demonstrated on \sco\ in the weak-field limit.}
{We propose three distinct statistical models, and use archival ESPaDOnS high-resolution Stokes $V$ observations to carry out the probabilistic magnetic inversion in closed form. The surface magnetic field is parameterised by a high-dimensional spherical-harmonic expansion.\looseness=-1}
{We provide mean magnetic field distributions along with the corresponding surface uncertainty maps for \sco. By comparing three different prior distributions over the latent variables in the spherical-harmonic decomposition, our results showcase the ZDI sensitivity to various hyperparameters. The mean magnetic field maps are qualitatively similar to previously published point estimates, but analysis of the magnetic energy distribution indicates high uncertainty and higher energy content at low angular degrees $l$.}
{Our results effectively demonstrate that, for stars in the weak-field regime, reliable uncertainty quantification of recovered magnetic field maps can be obtained in closed form with natural assumptions on the statistical model. Future work will explore extending this framework beyond the weak-field approximation and incorporating prior uncertainty over multiple stellar parameters in more complex magnetic inversion problems.}


\maketitle

\section{Introduction}\label{sec:introduction}

Magnetic fields are understood to play a significant role in the formation and evolution of stars. For example, magnetic interlocking of young stars with their accretion disks \cla{and winds} is the main mechanism behind stellar rotational braking \cla{\citep{Shu:1994,Bouvier:2014}}, and is therefore a key component of the early stellar evolution. At all stellar ages, a multitude of time-dependent surface phenomena and processes -- spots, flares, \cla{chromospheric emission} -- is believed to be driven by magnetic fields \cla{\citep{Reiners:2012}}. Extension of the surface field into the circumstellar environment -- the stellar magnetosphere -- shapes stellar winds, \cla{thereby governing the stellar angular momentum evolution \citep{Weber:1967,Matt:2008,Vidotto:2009},} and \cla{strongly impacts} the interaction of stars and nearby planets \cla{\citep{Lanza:2018,Strugarek:2016,Vidotto:2020}}.

Two basic types of magnetic fields are known to exist in stars \citep[e.g.][]{donati:2009}. On the one hand, all cool, low-mass stars, including the Sun and similar objects, continuously generate magnetic fields in their interiors through the so-called dynamo process. The resulting fields emerge at stellar surfaces in the form of evolving structures with moderate average strengths (10--100~G) and spatial scales ranging from a tiny fraction of the stellar surface (small-scale fields) all the way to the scales comparable to stellar radii (large-scale or global fields). On the other hand, the majority of hot, intermediate-mass and massive stars are not immediately recognised as magnetic. Only about 10\% \cla{or less, depending on the stellar mass,} of these objects possess detectable magnetic fields \cla{\citep{Grunhut:2017,Sikora:2019}}. \cla{Characteristics} of these fields are drastically different from those in cool stars. Magnetic fields of hot stars are topologically simple and strong ($\sim$\,kG), lack significant small-scale components, and are constant on the observable time scales.

Irrespective of the type of stellar magnetic field, information about its strength and surface geometry represents a critical stellar parameter highly sought after by theoretical and observational studies alike. Direct detection of stellar magnetic fields typically relies on the Zeeman effect in spectral lines \citep[e.g.][]{polarization:2004}. In the presence of a field, atomic energy levels split into magnetic sub-levels. This produces the corresponding splitting of spectral lines. Additionally, lines become polarised, with the direction of polarisation dependent on the field vector orientation and the amplitude of the polarisation signal linked to the field strength. This Zeeman-induced polarisation is particularly useful for detecting and characterising stellar magnetic fields.

Polarisation observations of stars are typically carried out using a combination of a high-resolution spectrograph and a circular polarimeter \citep[e.g.][]{donati:1997a}. Such instrumentation can measure the total intensity spectrum (Stokes $I$ parameter) together with the circular polarisation spectrum (Stokes $V$ parameter). The latter is sensitive to the line of sight component of the magnetic field. A complementary measurement of linear polarisation spectra (Stokes $Q$ and $U$ parameters), which provide information on the transverse field component, is also possible, but rarely accomplished due to a factor of 10 weaker linear polarisation signals \citep{wade:2000,kochukhov:2011,rosen:2015}.

Quantitative interpretation of the observed polarisation spectra of stars other than the Sun faces a significant challenge due to the unresolved nature of these observations. Polarisation signatures recorded by a distant observer represent an average of Stokes $V$ profiles originating from different locations on the visible stellar hemisphere, which makes it impossible to infer the magnetic field geometry from a single observation. Considerably more information is contained in a time sequence of polarisation observations of rotating magnetic stars. First, the rotational Doppler shift redistributes signals from different parts of the stellar surface within a spectral line profile, allowing one to relate the position in the line profile to the longitude on the stellar surface. Effectively, this makes a single line profile observation equivalent to a one-dimensional projection of the stellar surface map \citep{vogt:1987}. Second, the spectral line profile changes due to stellar rotation as different parts of the stellar surface come in and out of view. Combining these two effects in the inversion technique known as Doppler Imaging (DI) offers the possibility to reconstruct two-dimensional maps of the surfaces of unresolved distant stars by fitting a set of spectral line profiles recorded at different stellar rotation phases \citep[e.g.][]{kochukhov:2016a}.

DI was initially applied to map spots on different kinds of rapidly rotating stars with inhomogeneous surfaces using intensity spectra \citep{khokhlova:1986,vogt:1999}. The same inversion principles were subsequently extended to the problem of reconstructing the magnetic geometry of both cool stars with complex fields \citep{brown:1991,donati:1999} and hot stars with simple field topologies \citep{piskunov:2002,kochukhov:2002}. These early Zeeman-Doppler imaging (ZDI) studies approximated magnetic field distributions as a set of three independent images corresponding to the three magnetic field vector components, and employed either maximum entropy or Tikhonov regularisation to ensure convergence to a stable and unique solution. The majority of more recent ZDI applications use the spherical-harmonic decomposition to describe the stellar surface magnetic field \citep{donati:2006,kochukhov:2014,folsom:2018}. This approach ensures that the reconstructed magnetic field is solenoidal and provides a convenient framework to characterise poloidal/toroidal as well as axisymmetric/non-axisymmetric magnetic field components as a function of the spatial scale (angular degree $l$). In this case, spherical-harmonic coefficients, rather than the field vector component values at particular coordinates on the stellar surface, represent the free parameters of the inversion problem. Unless the stellar field is restricted to a very simple configuration (e.g. a pure dipole with only $l=1$ harmonics), regularisation is still required for stable recovery, due to the ill-posedness of the inverse problem.

Regardless of the specific implementation of ZDI, existing magnetic inversion methods suffer from one major shortcoming: it is difficult or outright impossible to provide realistic uncertainties of the derived magnetic field maps. Essentially all published mapping results represent point estimates, without an attempt to explore the full range of solutions compatible with observations. Due to the high dimensionality of the spectropolarimetric inversion problem and the complexity and computational intensity of the synthetic forward model, obtaining a probabilistic solution is challenging. Occasionally, uncertainties are assessed empirically by comparing magnetic field distributions computed from independent line profile observations \citep{kochukhov:2019,kochukhov:2022}. Some ZDI studies also considered formal uncertainties obtained from the diagonal of a Hessian matrix \citep{piskunov:2002}.  The former approach is, however, feasible only for strongly magnetic stars, and the latter yields biased uncertainties since parameter correlations are not considered. Additionally, several important stellar parameters (including rotational period, projected rotational velocity and inclination of the rotational axis) enter ZDI calculations. Each of these parameters is known to a limited precision, but their uncertainties are rarely propagated to magnetic maps \citep{petit:2008}. All these problems make it difficult to judge the reliability of published ZDI results, hindering their interpretation and wide utilisation.

In this paper, we address the problem of deriving realistic ZDI uncertainties by presenting a fully Bayesian framework for probabilistic ZDI. Previously, similar methods have only been applied to determine properties of stellar magnetic fields from circular spectropolarimetric observations under an oblique dipole model assumption \citep{petit:2012}, which is a restrictive assumption not suitable for many magnetic stars. Here, we present a general framework for probabilistic ZDI, using a probabilistic model based on a general spherical-harmonic field parameterisation. Compared to the dipolar modelling approach, our probabilistic ZDI framework applies to arbitrarily complex magnetic field geometries.
To present the proposed framework, we start with the comparatively simple problem of probabilistic ZDI reconstruction for a target star with fixed nuisance parameters and a weak magnetic field, which ensures a linear relation between the free parameters describing the magnetic map and the modelled polarisation signal. More specifically, we apply the Bayesian framework for uncertainty quantification of the surface magnetic field of the star \sco\ (HD\,149438, HR\,6165). In a future paper, we will extend our approach to stars exhibiting stronger magnetic fields, corresponding to a non-linear line profile response, and incorporate a Bayesian treatment of input stellar parameters.

Our target, \sco, is a bright, massive star with spectral type B0.2V. This star is located in the $\sim$\,10~Myr-old \citep{feiden:2016} Upper Sco stellar association. This star exhibits unusual characteristics, such as a slow rotation, a younger age than the surrounding stellar population, and a nitrogen excess \citep[e.g.][]{przybilla:2010,nieva:2014}, which distinguishes it from the other massive stars in Upper Sco. Even more remarkably, \citet{donati:2006} demonstrated through spectropolarimetric observations and ZDI modelling the presence of a relatively weak ($\le$\,600~G) and unusually complex magnetic field at the surface of that star. Unlike the majority of hot, magnetic stars, which possess a roughly dipolar (i.e. spherical-harmonic modes with $l=1$) field geometries \citep{shultz:2018}, \sco\ was found to exhibit a surface magnetic structure containing significant power in the harmonic modes of up to $l=5$, with the largest contribution coming from $l \in [3,4]$. 

These unusual chemical, rotational, and magnetic characteristics of \sco\ spurred a series of studies to explain this star as a product of a stellar merger \citep{nieva:2014,schneider:2016,keszthelyi:2021}. In particular, \citet{schneider:2019} presented 3D magnetohydrodynamical simulations tailored to \sco\ that demonstrated generation of a powerful magnetic field during a stellar merger event. This field decays quickly after the merger, settling into the stable non-axisymmetric configuration observed today \citep{schneider:2020,braithwaite:2008}. So far, \sco\ remains the only well-established magnetic star that is likely a merger product. Therefore, it plays the role of a key benchmark object for investigations into both binarity and magnetism of massive stars. At the same time, a follow-up ZDI analysis of \sco\ by \citet{kochukhov:2016}, using an extended observational dataset, suggested that the outcome of magnetic inversions based on circular polarisation spectra strongly depends on the assumptions of the magnetic field parameterisation. This non-uniqueness problem may arise in various applications of ZDI \citep[e.g.][]{donati:1997,kochukhov:2002}, but is particularly significant for \sco\ due to its field complexity and small amount of rotational Doppler broadening. Consequently, \citet{kochukhov:2016} obtained vastly different magnetic field geometries and strengths with equally good fits to available observations of \sco\ by using different, but equally plausible, harmonic parameterisations. These issues, and the significance of \sco\ for wider massive-star research, position this star as a particularly interesting target for probabilistic ZDI.  

This paper is organised in the following way: in Sect.~\ref{sec:obs_data}, the spectropolarimetric observations of \sco\ used for probabilistic spectropolarimetric inversion are introduced. \cla{Sect.~\ref{method} presents} a detailed description of the proposed Bayesian framework for probabilistic ZDI. We propose several probabilistic models, some of which aim to incorporate uncertainties in design parameters corresponding to hyperparameters in the regularisation function often used in conventional ZDI. 
Then, in Sect.~\ref{results}, we present summary statistics of the magnetic field map distributions obtained using our framework for probabilistic ZDI. Finally, in Sect.~\ref{discussion}, \cla{we discuss and summarise the results and conclusions}.

\section{\cla{Observational data and stellar parameters}}\label{sec:obs_data}

We use the same observations of \sco\ as analysed by \citet{kochukhov:2016}. The spectropolarimetric data available for this star comprises 49 circular polarisation observations collected in 2004--2009 with the ESPaDOnS instrument \citep{donati:2006b} at the 3.6m Canada-France-Hawaii telescope. The spectra, publicly available in the PolarBase archive\footnote{\url{http://polarbase.irap.omp.eu/}} \citep{petit:2014}, were processed with the least-squares deconvolution \citep[LSD,][]{kochukhov:2010} method, yielding high-quality mean Stokes $I$ and $V$ profiles. 

We refer the readers to \citet{kochukhov:2016} for further details on the data reduction and pre-processing. Following that study, as well as the earlier work by \citet{donati:2006}, we adopt a rotational period of $P_{\rm rot}=41.033$~d, a projected rotational velocity of \vsini\,=\,6~\kms\ and an inclination angle of $i=70\degr$ for the ZDI analysis, without accounting for possible errors in these parameters.

The observational data used in our work includes uncertainties for each velocity bin in the Stokes $I$ and $V$ parameter profiles. For data with high signal-to-noise ratio, as considered here, the errors are uncorrelated and follow a Gaussian distribution. For a detailed discussion of how these errors are derived by the data reduction pipelines from raw spectropolarimetric exposures and then propagated through the derivation of LSD profiles, we refer the readers to \citet{donati:1997a} and \citet{kochukhov:2010}. 

\section{\cla{Methods}}\label{method}

\subsection{Magnetic field map parameterisation}

There are multiple ways of parameterising the magnetic field distribution across the surface of a target star. Earlier ZDI inversions have adopted direct parameterisation techniques \citep[e.g.][]{brown:1991,kochukhov:2002}, in which each magnetic field vector component map is represented by the nodes of a discrete surface grid, and fitted to the data independently. Since then, alternative parameterisations of the surface magnetic field, employing spherical-harmonic expansions, have often been preferred \citep[e.g.][]{donati:2006,kochukhov:2014}. In contrast to the direct approach, this magnetic field parameterisation \cla{allows us to impose restrictions on the magnetic geometry to follow Maxwell's equations. Specifically, excluding the $l=0$ harmonic component ensures no net signed magnetic flux} through the closed stellar surface. There are several other appealing properties of the spherical-harmonic field parameterisation, see e.g. the discussion by \citet{kochukhov:2016}. 

In this work, we use the spherical-harmonic representation of the surface magnetic field defined as follows:
\begin{align}\label{eq:mag}
     B_r(\theta, \phi) &= -\sum^{l_\mathrm{max}}_{l=1} \sum^l_{m=-l} \alpha_{l,m} Y_{l, m}(\theta, \phi), \nonumber \\
     B_\theta(\theta, \phi) &= -\sum^{l_\mathrm{max}}_{l=1} \sum^l_{m=-l} \beta_{l,m} Z_{l, m}(\theta, \phi) + \gamma_{l, m} X_{l, m}(\theta, \phi), \\
     B_\phi(\theta, \phi) &= -\sum^{l_\mathrm{max}}_{l=1} \sum^l_{m=-l} \beta_{l,m} X_{l, m}(\theta, \phi) - \gamma_{l, m} Z_{l, m}(\theta, \phi). \nonumber
\end{align}
Here, $B_r(\theta, \phi), B_{\theta}(\theta, \phi)$, and $B_{\phi}(\theta, \phi)$ denote the surface magnetic field components in the radial, meridional, and azimuthal direction, respectively, with $\theta$ and $\phi$ \cla{corresponding to} the surface \cla{colatitude} and longitude. The spherical-harmonic functions \cla{and their derivatives} for each angular degree $l$ and azimuthal order $m$ are denoted by $Y_{l, m}(\theta, \phi), Z_{l, m}(\theta, \phi)$ and $X_{l, m}(\theta, \phi)$. Finally, the amplitudes of the spherical-harmonic modes, $\alpha_{l, m}$, $\beta_{l, m}$ and $\gamma_{l, m}$, delineate characteristics of the poloidal and toroidal magnetic field components. Together, these coefficients represent the free parameters in the spectropolarimetric inversion problem. See \citet{kochukhov:2014} for a more detailed description of the field parameterisation and relevant notations. \cla{One difference compared to \citet{kochukhov:2014} is that the constant $C_{l,m}$, defined by their Eq.~(7), was modified\footnote{\cla{This was accomplished by multiplying $C_{l,m}$ by $\sqrt{(l+1)/l}$ for $m=0$ and by $\sqrt{2(l+1)/l}$ for $m\ne0$ modes.}} here to ensure that equal values of the harmonic coefficients yield equal total magnetic energy, independent of $l$ and $m$.}

\subsection{Forward line profile model}\label{sec:forward}

In this study, we restrict our attention to stars for which the weak-field assumption \citep{polarization:2004} is a good approximation for the forward Stokes $I$ and $V$ profile model. This approach is typically valid for sub-kG fields studied with spectral lines at optical wavelengths and is applicable to a large group of stars, including solar-like stars, the majority of low-mass stars, and a few weak-field hot stars such as \sco. 

In practice, we follow \citet{petit:2012} in representing the local Stokes $I$ profile using a Gaussian function. \cla{Its} strength \cla{($d=0.115$)}, position \cla{(radial velocity shift $v_0=-0.48$~km\,s$^{-1}$)}, and width \cla{(FWHM\,=\,14.3~km\,s$^{-1}$ corresponding to $\sigma=6.1$~km\,s$^{-1}$)} \cla{are} adjusted to fit the observed LSD \cla{Stokes} $I$ spectrum. \cla{The adopted Gaussian line width includes the effect of intrinsic stellar line broadening as well as the smearing by the instrumental profile of ESPaDOnS ($R\approx68\,000$).}

The local Stokes $V$ profile is derived \cla{from the velocity} derivative of the Gaussian function
\begin{align}\label{eq:prf}
\cla{V(v)}~ & \cla{= -1.4 \times 10^{-3} \lambda_0 g_{\rm eff} B_\parallel \dfrac{\partial I}{\partial v}} \nonumber \\
& 
\cla{
=1.4\times10^{-3} \lambda_0 g_{\rm eff} B_\parallel \dfrac{d(v-v_0)}{\sigma^2} \exp{\left[{-\dfrac{(v-v_0)^2}{2\sigma^2}}\right]}.
}
\end{align}
Here, the \cla{central wavelength adopted for} the LSD profile is $\lambda_0=500$~nm and the effective Land\'e factor is $g_{\rm eff} = 1.2$ \citep[see][]{kochukhov:2016}. \cla{The quantity} $B_\parallel$ denotes the line of sight (longitudinal) component of the magnetic field \cla{(in kG)} at the $\theta,\phi$ location on the stellar surface\cla{, as seen by the observer at the rotational phase $t$, defined over the interval $[0, 1)$. This magnetic field projection can be calculated with}
\begin{align}\label{eq:bz}
\cla{B_\parallel =} & \cla{\phantom{-} \left[ \cos{\theta} \cos{i} + \sin{\theta} \sin{i} \cos{(\phi+2\pi t)} \right] B_r} \nonumber \\
 & \cla{- \left[ \sin{\theta} \cos{i} - \cos{\theta} \sin{i} \cos{(\phi+2\pi t)} \right] B_{\theta} } \\
 & \cla{- \sin{i} \sin{(\phi+2\pi t)} B_{\phi}.} \nonumber
\end{align}

To obtain the disc-integrated Stokes $I$ and $V$ profiles, we divide the stellar surface into $\approx$\,$10^4$ elements, \cla{pre-calculate the areas $S$ of these elements, and evaluate} the magnetic field using Eq.~(\ref{eq:mag}). \cla{Then, for each rotational phase $t$, we determine the cosine $\mu$ of the angle between the observer's line of sight and the surface normal (the limb angle) using}
\begin{equation}
\cla{\mu = \cos{\theta} \cos{i} + \sin{\theta} \sin{i} \cos{(\phi+2\pi t)} } 
\end{equation}
\cla{and locate visible surface elements with the condition $\mu>0$. Eq.~(\ref{eq:bz}) is then applied to determine $B_\parallel$ \cla{for visible surface elements}. Following this, the local Stokes $I$ and $V$ spectra are calculated as described above. This calculation is carried out on the velocity grid of observations, \revb{$v_{\rm obs}$}, Doppler-shifted according to the local \revb{component of the} projected rotational velocity, \revb{$v_{\rm e}\sin{i}$},}
\begin{equation}
\cla{ v = v_{\rm obs} - v_{\rm e}\sin{i} \sin{\theta} \sin{(\phi+2\pi t)}. } 
\end{equation}
\cla{Finally,} we add together all profiles from the visible stellar hemisphere using the product of the \cla{projected} surface area and a linear continuum limb-darkening function with the coefficient $u=0.3$ as \cla{an integration weight $W$}
\begin{equation}
\cla{ W = S \mu (1-u+u \mu) }
\end{equation}

\cla{Due to the weak-field assumption expressed by Eq.~(\ref{eq:prf}), this sequence of operations represents a linear transformation from a set of the spherical harmonic coefficients, $\alpha$, $\beta$ and $\gamma$, to the phase-dependent disk-integrated Stokes $V$ profiles. We take advantage of this linearity of the problem in this paper. At the same time, we anticipate a generalisation to the strong-field situation in future studies by replacing Eq.~(\ref{eq:prf}) with another relation, not limited to weak fields.}

While we have used the observed Stokes $I$ LSD profiles to adjust the Gaussian function parameters and determine a radial velocity offset, the magnetic field does not impact Stokes $I$ in the weak-field limit. Consequently, the classical and probabilistic ZDI analyses presented in this paper consider only the Stokes $V$ observations.

\subsection{Standard ZDI}\label{sec:ZDI}

We define the vector $\mathbf{z}~=~(\alpha_{l, m}, \beta_{l, m}, \gamma_{l, m})$, where $m~\in~\{-l, -l+1, \ldots, l-1, l\}$ for each $l \in \{1, 2, \ldots, l_\mathrm{max}\}$, as a collection of the amplitudes of the spherical-harmonic modes, $\alpha_{l, m}, \beta_{l, m}$, and $\gamma_{l, m}$. We truncate the expansion by $l_{\mathrm{max}} = 10$, corresponding to 360 free parameters in the resulting magnetic field
parameterisation. In a nutshell, standard ZDI amounts to solving the weighted least-squares (LS) problem 
\begin{equation} 
    \hat{\mathbf{z}} = \argmin_{\mathbf{z}} \left( \lVert \Lambda^{\frac{1}{2}} (\mathbf{y} - \mathbf{f}(\mathbf{z})) \rVert_2 ^2 + \eta \, r(\mathbf{z}) \right),\label{eq:opt}
\end{equation}
with respect to the spherical-harmonic coefficients $\mathbf{z}$. In the above formulation, $\mathbf{f}(\mathbf{z})$ denotes the forward mapping function (corresponding to the computations described in Sect.~\ref{sec:forward}) between $\mathbf{z}$ and the observed Stokes~$V$ spectra denoted $\mathbf{y}$. Moreover, $\Lambda$ denotes the inverse covariance matrix of the measurement noise (i.e. the precision matrix). Another important component is the regulariser
\begin{equation} 
    r(\mathbf{z})~= \sum_{l=1}^{l_{\mathrm{max}}} \sum_{m=-l}^{l} l^2 (\alpha_{l, m}^2 + \beta_{l, m}^2 + \gamma_{l, m}^2),\label{eq:opt_regularizer}
\end{equation}
with regularisation strength $\eta$.
The regulariser ensures that, despite the ill-posedness of the Stokes $V$-only ZDI inversion \citep{piskunov:2002}, a unique solution exists, that avoids the introduction of higher-order modes unless the added flexibility is motivated by the data. The regularisation structure presented in Eq.~\eqref{eq:opt_regularizer}, where the regularisation objective is to minimize the $l^2$-weighted magnetic energies of the spherical-harmonic coefficients, is commonly used in some ZDI studies \citep[e.g.][]{Hussain:2000,kochukhov:2014,Kochukhov2016} while others adopt the $l$-weight for this regularisation \citep{Morin:2008,Morin:2010}.

The first part of the optimisation objective in Eq.~\eqref{eq:opt} is hereon referred to as the weighted LS objective, and the second part is referred to as the regularisation objective. Since $\eta$ determines the trade-off between these two optimisation objectives, the choice of $\eta$ has a significant impact on the properties of the solution. Larger $\eta$ may result in information loss by favouring the generation of smoother magnetic field maps. On the other hand, small $\eta$ may introduce spurious high-order modes, resulting in magnetic field maps containing unjustified surface details. To choose the regularisation parameter, we follow the empirical approach of choosing $\eta$ such that the quotient between the weighted LS objective value and the regularisation objective value at the optimum is between 2 and 4. To briefly summarise the underlying motivation, this region defines a breaking point after which the fit quality improves slowly as the regularisation strength decreases. This approach to choosing the regularisation parameter is discussed in more detail by \citet{Kochukhov2017}.

This classical ZDI framework views the model parameters $\mathbf{z}$ as fixed quantities to infer. To express uncertainty about the resulting point estimates $\hat{\mathbf{z}}$ of these quantities, the distribution of possible datasets can be considered. However, data collection in the context of ZDI is very costly and stellar magnetic fields often evolve between different observing runs, rendering this approach impractical. Moreover, due to the ill-posedness of the problem, formal error bars on $\hat{\mathbf{z}}$ become strongly dependent on the regularisation strength $\eta$. For these reasons, and due to computational limitations of previous ZDI approaches, such error bars have generally not been presented in previous ZDI studies. In the following section, we describe a procedure for quantifying uncertainty in the parameters $\mathbf{z}$ according to the Bayesian viewpoint.

\subsection{Probabilistic ZDI}

In this section, we propose several statistical models to extend the standard ZDI framework, discussed in Sect.~\ref{sec:ZDI}, into a fully Bayesian setting. The Bayesian setting \citep[e.g.][]{bayesian_data_analysis} provides a framework for updating prior beliefs about the uncertainty in the spherical-harmonic coefficients, represented as a probability distribution over $\mathbf{z}$, as new data becomes available. To this end, we treat the spherical-harmonic coefficients $\mathbf{z}$ as latent random variables. Given a set of spectropolarimetric observations $\mathbf{y}$, the resulting posterior distribution, $p(\mathbf{z}|\mathbf{y})$, captures our informed state of knowledge about the parameters $\mathbf{z}$ conditioned on $\mathbf{y}$. Conveniently, this framework brings inherent regularisation into the inversion process; an especially attractive property in light of the ill-posedness of the problem at hand. We begin by constructing a probabilistic model and describing how to calculate the posterior distribution in the context of probabilistic ZDI. At the end of the section, we dive deeper into the choice of prior distribution, aiming to capture uncertainty propagated through its hyperparameters.  

\subsubsection{Model formulation}\label{model_formulation}

To model the likelihood, the observational noise is assumed to be Gaussian with diagonal covariance (see Sect. \ref{sec:obs_data}), such that $p(\mathbf{y}|\mathbf{z}) = \mathcal{N}(\mathbf{y}; \mathbf{f}(\mathbf{z}), \Lambda^{-1})$\footnote{\cla{The notation $p(\mathbf{x}) = \mathcal{N}(\mathbf{x}; \mathbf{\mu}, \Sigma)$ represents a Gaussian distribution over the random variable $\mathbf{x}$, with mean vector $\mathbf{\mu}$ and covariance matrix $\Sigma$.}}. Then, given a prior distribution $p(\mathbf{z})$, the posterior distribution $p(\mathbf{z}|\mathbf{y})$ can be obtained following Bayes' theorem,  according to
\begin{equation}\label{eq:bayes_rule}
p(\mathbf{z}|\mathbf{y}) = \frac{p(\mathbf{y}|\mathbf{z})p(\mathbf{z})}{p(\mathbf{y})}.
\end{equation}
\cla{Given $p(\mathbf{z}|\mathbf{y})$, the posterior magnetic field distribution can then be obtained by applying the transformation in Eq. \eqref{eq:mag}}.
In general, exact determination of the posterior distribution according to Eq.~\eqref{eq:bayes_rule} is challenging since the marginal likelihood $p(\mathbf{y})$ is often intractable. When the posterior distribution has been inferred, the predictive distribution over \cla{the modelled Stokes V profiles} $\mathbf{y_*}$ can be obtained by marginalisation:
\begin{equation}\label{eq:pred_dist}
p(\mathbf{y_*} | \mathbf{y}) = \int p(\mathbf{y_*} | \mathbf{z}) p(\mathbf{z} | \mathbf{y}) d\mathbf{z}.
\end{equation}

We model the prior distribution by a Gaussian distribution, $p(\mathbf{z}) = \mathcal{N}(\mathbf{z};\mathbf{0}, \mathbf{\Omega}^{-1})$, where $\mathbf{\Omega}^{-1}$ is a diagonal covariance matrix. Despite the Gaussian likelihood and Gaussian prior distribution, Eq.~\eqref{eq:bayes_rule} is in general intractable if the forward mapping function $\mathbf{f}(\mathbf{z})$ is non-linear.

In the initial analysis, we use the parameters $\Omega_{i,i} = \eta l_i^2$ in the marginal prior distribution for each spherical-harmonic coefficient $z_i$. While it is difficult to construct a fully physics-informed prior distribution over the spherical-harmonic coefficients, this choice of prior distribution captures our prior belief that spherical-harmonic coefficients of high degree $l$ are less prevalent in the solution. In addition, the prior variance adds implicit regularisation to the solution, and with the proposed likelihood model and prior distribution, the resulting maximum a posteriori \cla{probability (MAP)} estimate, $\hat{\mathbf{z}}_{\mathrm{MAP}} = \argmax_{\mathbf{z}} \log p(\mathbf{z}|\mathbf{y})$, can be shown to solve the regularised weighted LS problem arising in the standard ZDI formulation given by Eq.~\eqref{eq:opt}. In that sense, the proposed probabilistic model allows us to generalise the regularisation structure often used in standard ZDI inversion (see Eq.~\eqref{eq:opt_regularizer}), with the hyperparameter $\eta$ in the prior distribution corresponding to the regularisation strength in the standard ZDI formulation. This means that, if the mean of the full posterior distribution coincides with the \cla{MAP} estimate, the Bayesian formulation results in an uncertainty quantification centred around the point estimate obtained from standard ZDI. \looseness=-1

\subsubsection{Posterior inference}\label{sec:post_inf}

In the weak-field regime, adopted in this paper, the forward model can be approximated by a function linear in the parameters $\mathbf{z}$ such that $\mathbf{f}(\mathbf{z}) = \mathbf{A}\mathbf{z}$, where $\mathbf{A}$ is the transformation matrix. \cla{This linearity follows from the set of equations presented in Sect. \ref{sec:forward}}. In this case, the Gaussian prior in our probabilistic model is conjugate to the likelihood, and a closed-form expression for the posterior distribution exists \citep[e.g.][]{pml1Book, bishop2007, Sjlund2017}. 
In fact, \cla{following the derivation in \citet[][Sect. 3.3.2]{pml1Book}}, it can be shown that the posterior distribution is also a Gaussian $p(\mathbf{z}|\mathbf{y})~=~\mathcal{N}(\mathbf{z}; \mathbf{\mathbf{\mu}}, \Sigma)$ with $\Sigma^{-1} = \mathbf{\Omega} + \mathbf{A}^T\Lambda\mathbf{A}$ and $\mathbf{\mu} = \Sigma \mathbf{A}^T\Lambda\mathbf{y}$.
Since $p(\mathbf{z}|\mathbf{y})$ is Gaussian and, according to Eq.~\eqref{eq:mag}, the magnetic field $B$ is a linear function of the spherical-harmonic coefficients $\mathbf{z}$, the posterior magnetic field distribution is also Gaussian. \cla{This distribution} can be expressed in closed form \cla{according to the linear transformation theorem for the multivariate Gaussian distribution}: 
if $\mathbf{z} \sim p(\mathbf{z}|\mathbf{y}) = \mathcal{N}(\mathbf{z}; \mathbf{\mathbf{\mu}}, \Sigma)$, then $B_r = \Tilde{B_r}\mathbf{z} \sim \mathcal{N}(\Tilde{B_r}\mathbf{\mu}, \Tilde{B_r}\Sigma{\Tilde{B_r}}^T), B_{\theta} = \Tilde{B_{\theta}}\mathbf{z} \sim \mathcal{N}(\Tilde{B_{\theta}}\mathbf{\mu}, \Tilde{B_{\theta}}\Sigma{\Tilde{B_{\theta}}}^T)$ and $B_{\phi} = \Tilde{B_{\phi}}\mathbf{z} \sim \mathcal{N}(\Tilde{B_{\phi}}\mathbf{\mu}, \Tilde{B_{\phi}}\Sigma{\Tilde{B_{\phi}}}^T)$. \cla{Here,} $\Tilde{B_r}, \Tilde{B_{\theta}}$ and $\Tilde{B_{\phi}}$ are defined according to the linear transformations in Eq.~\eqref{eq:mag}.

In this setting, the predictive distribution\cla{, see Eq. \eqref{eq:pred_dist},} is also available in closed form. \cla{We use $\mathbf{y}_*$ to denote the vector representing the modelled LSD Stokes V profiles, whereas $\mathbf{y}$ continues to denote the specific observations used to fit the probabilistic model. The corresponding transformation matrix used to make predictions is then denoted by $A^*$. This transformation matrix can be different from the transformation matrix $A$ used to fit the model, for example if we wish to increase the discretisation of the predicted Stokes $V$ profiles compared to the observations $\mathbf{y}$.} Given the transformation matrix $A^*$, the predictive distribution over $\mathbf{y_*}$ is then given by $p(\mathbf{y_*}|\mathbf{y}) = \mathcal{N}(\mathbf{y_*}; \mathbf{\mathbf{\mu_*}}, \Sigma_*)$, where $\mu_* = A^*\mu$ and $\Sigma_*~=~A^*\Sigma{(A^*)}^T + {\Lambda_{\cla{*}}}^{-1}$. \cla{Here, ${\Lambda_{*}}^{-1}$ is a diagonal matrix containing the estimated observational noise corresponding to $y_*$. This result follows from Eq. 2.115 in \citet{bishop2007}}.

\subsubsection{Prior hyperparameter selection}\label{sec:extended}

In the statistical model described in the previous subsections, a specific set of hyperparameters $\Omega_{i,i}$ is assumed to parameterise the variance in the prior distribution. In our initial analysis, we use the hyperparameters $\Omega_{i,i} = \eta l_i^2$ with $\eta=100$. Here, $\eta$ is determined according to the empirical approach described in Sect.~\ref{sec:ZDI}, motivated by the connection to the regularisation strength in the standard ZDI formulation. However, there is no consensus regarding the choice of regularisation in existing ZDI literature. Other authors favour a regularisation function that, in the probabilistic model formulation, corresponds to the hyperparameters $\Omega_{i,i} = \eta l_i$ (without the square) in the prior variance \citep{Morin:2008}. Previous studies also adopt different approaches to choosing the hyperparameter $\eta$ or use an entirely different form of regularisation altogether \citep{folsom:2018}. Even using a single approach, like the empirical approach described in Sect.~\ref{sec:ZDI}, there is an inherent uncertainty in the choice of $\eta$. Since the proposed framework for probabilistic ZDI quantifies uncertainty within the chosen model class, the uncertainty estimate can become more reliable by broadening the class of models under consideration. This can be achieved by modifying the prior distribution. In Sect. \ref{sec:empirical_bayes}-\ref{sec:mixture_priors}, we describe three ways of selecting hyperparameters in the prior distribution; either as a point estimate or by incorporating prior uncertainty over these parameters.

\subsubsection{Empirical Bayes}
\label{sec:empirical_bayes}

One method for determining hyperparameters in the prior distribution of a probabilistic model is called \emph{empirical Bayes}, or \emph{evidence maximisation} \citep[e.g.][]{bishop2007}. With this approach, \cla{a} point \cla{estimate} $\hat{\eta}$ of the hyperparameter $\eta$ is determined by maximising the marginal likelihood $~p(\mathbf{y}|\eta)=\int p(\mathbf{y}|\mathbf{z})p(\mathbf{z}|\eta)d\mathbf{z}$, i.e.
\begin{equation}\label{eq:evidence_maximization}
\hat{\eta} = \argmax_{\eta} \log p(\mathbf{y}|\eta).
\end{equation}
\cla{To make the dependence on the hyperparameter $\eta$ explicit here, 
$p(\mathbf{z}|\eta)$ denotes the prior distribution over the model parameters $\mathbf{z}$, parameterised by $\eta$.} The marginal likelihood \cla{quantifies how well the model using the hyperparameter $\eta$, with latent variables $\mathbf{z}$ governed by the prior distribution $p(\mathbf{z}|\eta)$, explains the observed data $\mathbf{y}$.}
The empirical Bayes hyperparameter estimate can then be viewed as a maximum likelihood estimate given a likelihood function defined over a specified model space, from which the random variables $\mathbf{z}$ have been marginalised out. 
In relation to the fully Bayesian approach to hyperparameter optimisation, we can view the empirical Bayes approach as an approximation of the posterior distribution $p(\eta|\mathbf{y})$ by a point function at its mode\cla{. This point estimate is} equivalent to the point estimate obtained assuming an improper prior distribution according to
\begin{equation}\label{eq:evidence_maximization2}
\hat{\eta} = \argmax_{\eta} p(\eta|\mathbf{y}) =  \argmax_{\eta} p(\mathbf{y}|\eta) p(\eta) = \argmax_{\eta} p(\mathbf{y}|\eta), 
\end{equation}
where, in the last step, we let $p(\eta) = 1$. As we can see, the point estimates in Eq.~\eqref{eq:evidence_maximization} and Eq.~\eqref{eq:evidence_maximization2} are identical.

We determine the hyperparameter $\eta$ given the probabilistic model described in Sect.~\ref{model_formulation} by solving the maximisation problem in Eq.~\eqref{eq:evidence_maximization}. Given the conjugacy between the likelihood and our choice of prior, a closed-form expression for the marginal likelihood exists in this case, namely $p(\mathbf{y}|\eta)~=~\mathcal{N}(\mathbf{y};~\mathbf{0}, \Lambda^{-1}~+~\mathbf{A}\mathbf{\Omega}^{-1}(\eta)\mathbf{A}^T)$. \cla{This result is derived using Eq. 2.115 in \citet{bishop2007}.}

\begin{figure}[t]
    \centering
    \includegraphics[width=\columnwidth]{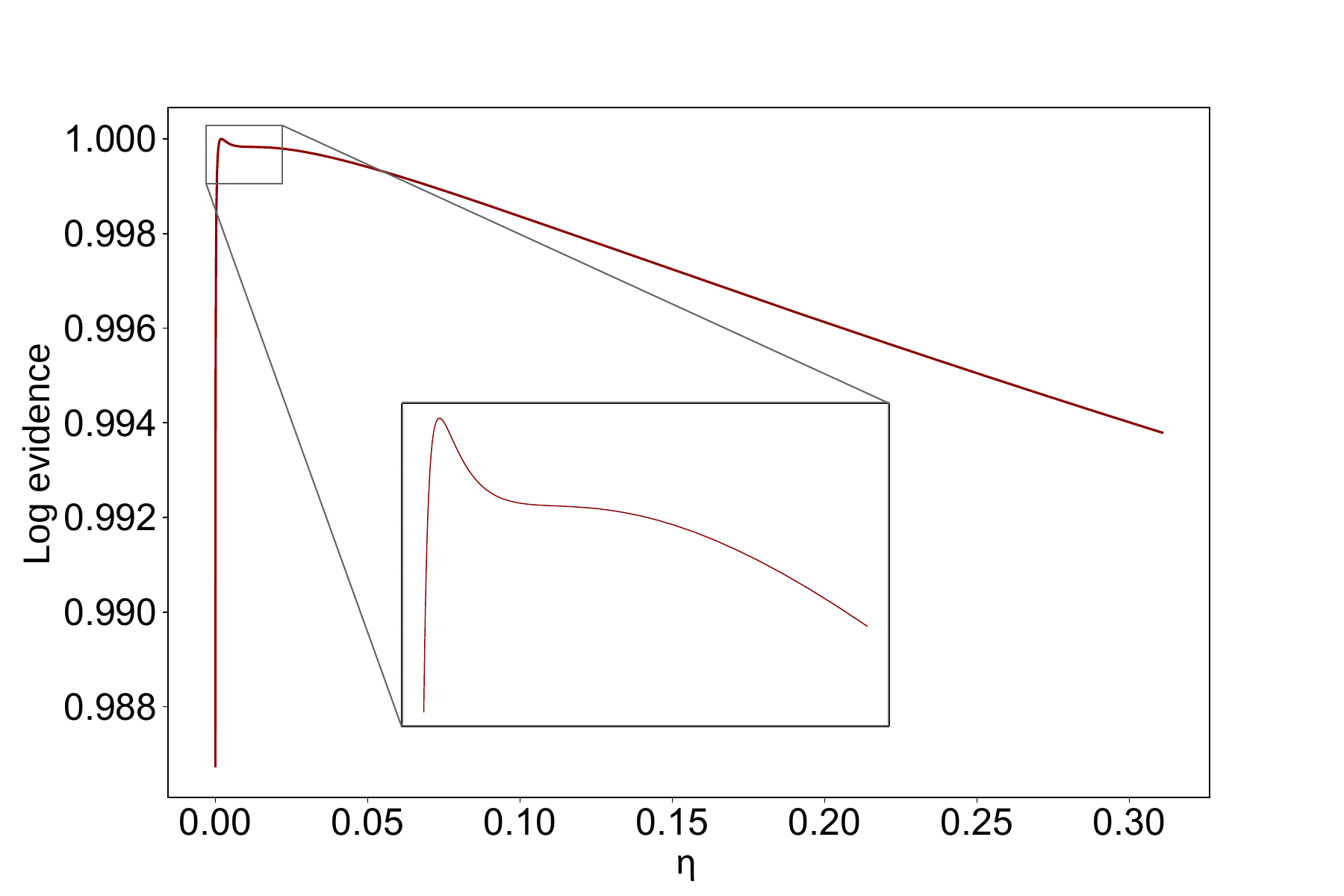}
    \caption{The unnormalised log evidence as a function of $\eta$ on a region close to $\eta=0$. The log evidence is maximised for $\eta \approx 0.002$. As $\eta$ grows outside of the displayed region, the log evidence continues to decrease.}
    \label{fig:img_emp_bayes}
\end{figure}

Using a prior distribution with parameters $\Omega_{i,i} = \eta l_i^2$ in the statistical model described in Sect.~\ref{model_formulation}, the empirical Bayes estimate $\hat{\eta}$ is in the order of $10^{-3}$, yielding a quotient exceeding a magnitude of $15$ between the weighted LS objective value and the regularisation objective value. Fig.~\ref{fig:img_emp_bayes} shows the log evidence as a function of $\eta$ near the maximum. 
In other words, the resulting quotient is far outside the interval $[2,4]$ at the mean of the posterior distribution $p(\mathbf{z}|\mathbf{y})$. The resulting magnetic field strength becomes too large, violating the weak-field assumption on the magnetic field used to model the likelihood. Thus, estimating $\hat{\eta}$ using empirical Bayes gives a magnetic field with properties inconsistent with the underlying physical assumptions even at the mean of the posterior distribution, suggesting that the marginal likelihood, and by extension empirical Bayes, is not useful in this particular context. \cla{We conclude that maximum likelihood estimation of this hyperparameter is ill-posed in this case, and we observe that the marginal likelihood is sensitive to small variations in $\eta$. Analysing the well-posedness of maximum likelihood estimation in the regression setting is non-trivial~\citep[see, e.g.,][for a deeper exposition]{karvonen2023}.
Consequently, maximum likelihood estimates in such settings may not reliably respect underlying assumptions.}

\subsubsection{Hierarchical model} \label{hyperpriors_section}

In a fully Bayesian setting, we treat hyperparameters in the prior distribution $p(\mathbf{z})$ as latent random variables in an extended probabilistic model. Let's consider the case in which we treat the prior hyperparameter $\eta$, corresponding to the regularisation strength in the standard ZDI framework, as a random variable. Prior knowledge of the distribution over $\eta$ can then be incorporated into the joint prior distribution, $p(\mathbf{z}, \eta$). The posterior distribution $p(\mathbf{z}, \eta|\mathbf{y})$ is then
\begin{equation}\label{eq:bayes_hyperprior}
p(\mathbf{z}, \eta|\mathbf{y}) = \frac{p(\mathbf{y}|\mathbf{z})p(\mathbf{z}, \eta)}{p(\mathbf{y})} = \frac{p(\mathbf{y}|\mathbf{z})p(\mathbf{z}|\eta)p(\eta)}{p(\mathbf{y})}.
\end{equation}
Here, $p(\eta)$ is referred to as the \emph{hyperprior distribution}. Treating hyperparameters of the prior distribution as random variables results in a so-called \emph{hierarchical probabilistic model}. The sought posterior distribution is then given by marginalising over $\eta$ according to
\begin{equation}\label{eq:bayes_marginal_posterior}
p(\mathbf{z}|\mathbf{y}) = \int p(\mathbf{z}, \eta|\mathbf{y}) d\eta,
\end{equation}
which is equivalent to assuming the prior distribution 
\begin{equation}\label{eq:bayes_marginal_posterior2}
p(\mathbf{z}) = \int p(\mathbf{z}, \eta) d\eta = \int p(\mathbf{z}|\eta)p(\eta) d\eta
\end{equation}
in Eq.~\eqref{eq:bayes_rule}. Defining a suitable prior distribution $p(\eta)$ can be challenging, and this distribution will also be defined in terms of its own set of hyperparameters that need to be chosen. 
We can extend the probabilistic model with another hierarchical layer by treating the hyperparameters in the hyperprior distribution as random variables, but the hyperparameters in the final hierarchical layer eventually need to be fixed. The limiting distribution, however, often becomes invariant to the specific form of the hyperprior distributions, and the specific hyperparameters in the hyperprior distributions therefore become less and less important as the number of hierarchical layers grows, as long as the distributions are sufficiently broad \citep[see, e.g.][]{roberts2001infinite}.

\begin{figure}[t]
    \centering
    \includegraphics[width=\columnwidth]{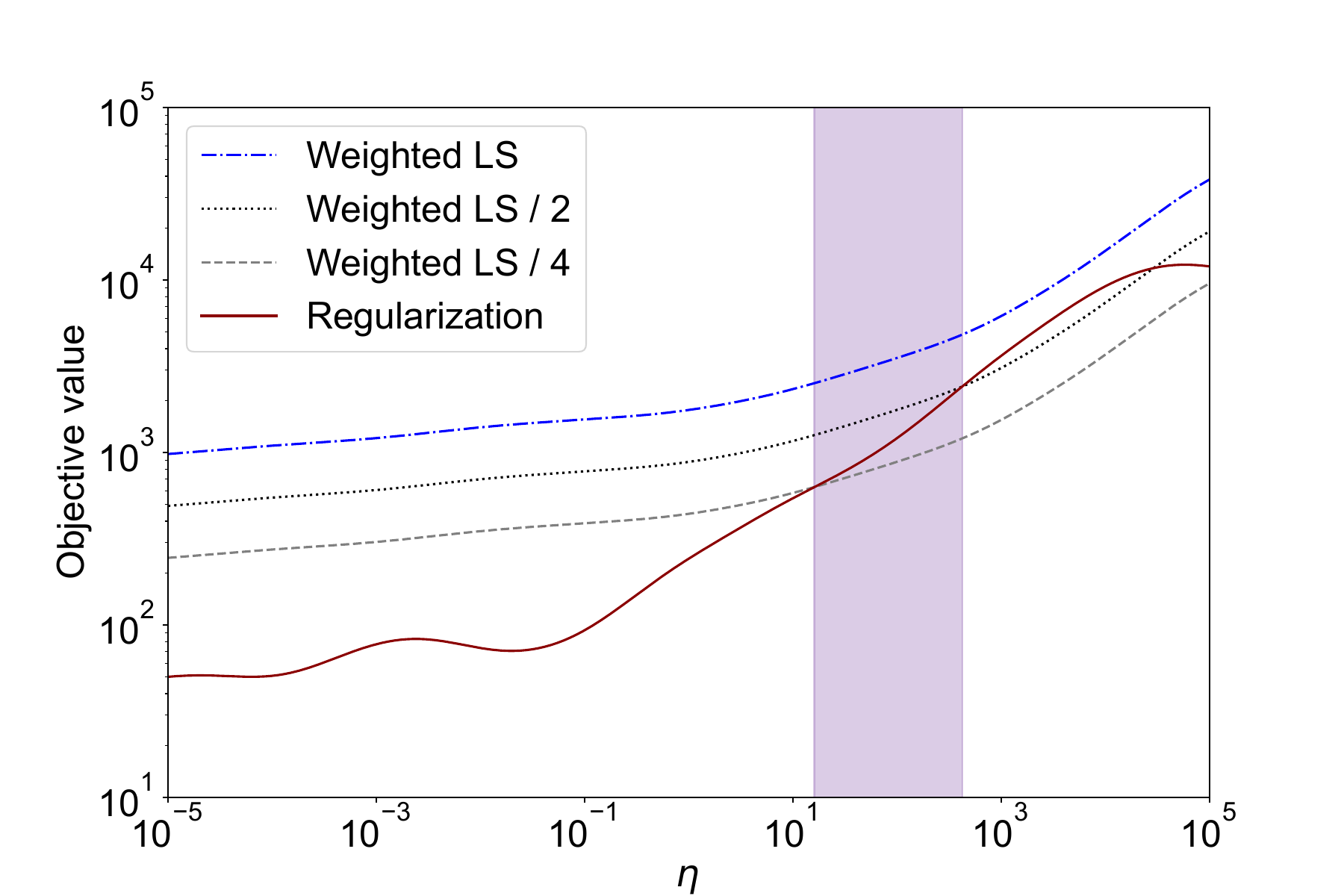}
    \caption{The trade-off between the least-squares fit and the 
    regularisation objective as a function of $\eta$. The region where $\eta \in [16, 421]$ is shaded in purple.}
    \label{fig:img_reg}
\end{figure}

We restrict our attention to the one-layer hierarchical model in Eq.~\eqref{eq:bayes_hyperprior}, and begin by modelling the hyperprior distribution $p(\eta)$. \cla{Its exact form is a modelling choice, and one possible approach is to let the hyperprior distribution} be informed by the empirical approach to selecting $\eta$ described in Sect.~\ref{sec:ZDI}. Recall that this approach suggests that a reasonable point estimate of $\eta$ can be obtained in the range within which the quotient between the weighted LS objective value and the regularisation objective value is between 2 and 4 using standard ZDI. For \sco, this region is illustrated in Fig.~\ref{fig:img_reg}\revb{. The figure shows the two components of the objective function in Eq. \eqref{eq:opt}, i.e. the weighted LS objective and the regularisation objective, evaluated at the solution of the optimisation problem for a range of $\eta-$values. Specifically, for each $\eta$, we solve the optimisation problem given in Eq. \eqref{eq:opt}, and plot the resulting weighted LS (data fidelity) terms $\lVert \Lambda^{\frac{1}{2}} (\mathbf{y} - \mathbf{f}(\mathbf{\hat{z}})) \rVert_2 ^2$ and regularisation terms $\eta \, r(\mathbf{\hat{z}})$ at the solutions $\mathbf{\hat{z}}$ as a function of $\eta$.} The region over $\eta$ where the regularisation term is between 2 and 4 times smaller than the LS objective is shaded in purple, and corresponds to $\eta \in [16, 421]$. \revb{As can be seen, the improvement in fit quality (reflected by a decreasing weighted LS objective value) begins to slow down in the vicinity of this region, as $\eta$ decreases from the right in the plot. Moreover, the difference in magnitude between the weighted LS objective value and the regularisation term increases.}
\revb{Informed by this approach to hyperparameter selection in standard ZDI, we can model $p(\eta)$ as a distribution with non-zero probability on the interval $\eta \in [16, 421]$}. 

However, exactly inferring the posterior distribution as defined in Eq.~\eqref{eq:bayes_hyperprior} is generally intractable, as it is rare to have conjugacy in a hierarchical model even with a single hyperprior distribution. Thus, adopting a hierarchical modelling approach to incorporating uncertainty over the hyperparameters in our prior distribution requires approximate solution methods, such as Markov chain Monte Carlo (MCMC) methods \citep[see, e.g.][]{meyn_tweedie2009, cotter2013}.
While MCMC methods can provide statistically exact sample approximations of the posterior distribution, the computational demand can be prohibitive, especially in high-dimensional settings such as those arising from high-order spherical-harmonic magnetic field parameterisations. This challenge is further compounded when the forward model is even mildly computationally intensive \citep[][]{brooks2011handbook}.
Since we are operating in the weak-field regime in our analysis of \sco, under which our probabilistic model formulation conveniently allows for a closed-form expression of the posterior distribution in Eq.~\eqref{eq:bayes_rule}, we are interested in finding an alternative probabilistic model incorporating uncertainty in the hyperparameter $\eta$, allowing for a closed-form expression of the posterior distribution in Eq.~\eqref{eq:bayes_marginal_posterior}. This can be achieved by approximating the continuous hyperprior distribution over $\eta$ by a discrete distribution, obtained by evenly discretising the interval of non-zero hyperprior probability, i.e. $\eta \in [16, 421]$, in $C$ components, and assigning each component a prior probability, or weight, $\omega_c$. This is equivalent to modelling $p(\mathbf{z})$ as a \emph{mixture prior} to incorporate our prior belief about $\eta$ directly into the prior distribution, in place of explicitly extending the model with a hyperprior distribution $p(\eta)$. A detailed account of this approach is provided in the following section.

\subsubsection{Mixture priors}
\label{sec:mixture_priors}
In this section, we extend the probabilistic model in two ways compared to the initial model described in Sect.~\ref{model_formulation}, both of which adjusts the prior distribution to incorporate prior uncertainty over hyperparameters in the prior variance using mixture priors. A mixture prior can be constructed according to
\begin{equation}\label{eq:prior}
    p(\mathbf{z}) = \sum_{c=1}^{C} \omega_c p_c(\mathbf{z}),
\end{equation}
where $C$ is the number of mixture components and $p_c(\mathbf{z})$ denotes the $c-$th prior component. In this formulation, each prior component is a properly normalised probability density function and $\sum_{c=1}^{C} \omega_c = 1$, such that the mixture prior $p(\mathbf{z})$ is also properly normalised. With this notation, the posterior distribution can be expressed as
\begin{equation}\label{eq:post}
    p(\mathbf{z}|\mathbf{y}) \propto \sum_{c=1}^{C} \omega_c p_c(\mathbf{z})p(\mathbf{y}|\mathbf{z}),
\end{equation}
where $p(\mathbf{y}|\mathbf{z})$ is the likelihood. Now, we define the marginal likelihood for each component $c$ according to ~$p_c(\mathbf{y})=\int p_c(\mathbf{z}) p(\mathbf{y}|\mathbf{z}) d\mathbf{z}$. We can then rewrite Eq.~\eqref{eq:post} according to
\begin{equation}\label{eq:post2}
    p(\mathbf{z}|\mathbf{y}) \propto \sum_{c=1}^{C} \omega_c p_c(\mathbf{y}) \frac{p_c(\mathbf{z})p(\mathbf{y}|\mathbf{z})}{p_c(\mathbf{y})},
\end{equation}
where $p_c(\mathbf{z})p(\mathbf{y}|\mathbf{z})/{p_c(\mathbf{y})}$ is the posterior distribution obtained using the prior distribution component $p_c(\mathbf{z})$. Thus, using a mixture prior \cla{in} the form of Eq.~\eqref{eq:prior}, the posterior distribution is a mixture of the posterior distributions obtained from each component in the mixture prior distribution. The unnormalised weights in the posterior mixture are given by $\omega_c p_c(\mathbf{y})$ for each component $c$, where $\omega_c$ are the weights of the components in the mixture prior distribution. Properly normalised, the mixture posterior distribution becomes
\begin{equation}\label{eq:post3}
    p(\mathbf{z}|\mathbf{y}) = \left({\sum_{c=1}^{C} \omega_c p_c(\mathbf{y})}\right)^{-1} \sum_{c=1}^{C} \omega_c p_c(\mathbf{y}) \frac{p_c(\mathbf{z})p(\mathbf{y}|\mathbf{z})}{p_c(\mathbf{y})}.
\end{equation}
 
 The connection between this model and the hierarchical model with a prior defined according to Eq.~\eqref{eq:bayes_marginal_posterior2} is now clear. The mixture prior can be viewed as a discretisation of the integral arising when marginalising out $\eta$, with the mixture weights $\omega_c$ forming a discrete hyperprior distribution $p(\eta)$. Equal weights for each prior component then correspond\cla{s} to a discrete uniform hyperprior distribution over $\eta$. We can also view each mixture component as a separate \emph{model}, and interpret the prior weights as an expression of our prior belief in each model. With this view, our approach to incorporating uncertainty of prior hyperparameters into the probabilistic model is equivalent to \emph{Bayesian model averaging} (see, e.g \citet{HOET1999}). 

 We begin by using a mixture prior to capture our prior uncertainty over the hyperparameter $\eta$ in the prior variance term. The contribution of each component $c$ to the prior distribution is modelled as a Gaussian according to $p_c(\mathbf{z}) = \mathcal{N}(\mathbf{0}, \mathbf{{\Omega}^{-1}_c})$, where $\mathbf{{\Omega}^{-1}_c}$ is a diagonal covariance matrix. We use the parameter ${\Omega_{c}}_{i,i} = \eta_c l_i^2$ in the marginal prior distribution for each spherical-harmonic coefficient $z_i$, but now the prior variance depends on $c$. 
 We use a mixture prior with $C = 1000$ components, where the parameters $\eta_c \in [\eta_1,\eta_C]$ are evenly distributed on the interval $[16, 421]$, according to the earlier motivation.
 
 It remains to determine the weights $\omega_c \in [\omega_1, \omega_C]$. According to Eq.~\eqref{eq:post3}, the weights in the mixture posterior distribution are scaled by the corresponding marginal likelihoods $p_c(\mathbf{y})$ compared to the weights in the mixture prior. The component-wise marginal likelihood is heavily dependent on $\eta$, and varies significantly in magnitude depending on the value of $\eta_c$. In fact, using equal prior weights $\omega_c$ for all components $c \in [1,1000]$ results in negligible weights for all components in the mixture posterior, except the component $c=1$. This prior component uses $\eta_1$, the lowest possible value of $\eta$ within our prior interval of non-zero probability. In light of this reflection, we note that the use of a mixture prior with equal weights in this case \cla{gives a result that is approximately} equivalent to selecting $\eta$ using \emph{empirical Bayes} as discussed \revb{in Sect. \ref{sec:empirical_bayes}}, \cla{with an additional requirement} constraining $\eta$ to the region $[16,421]$ when solving the optimisation problem in Eq.~\eqref{eq:evidence_maximization}. 
 \cla{Thus,} the choice of using uniform weights may not provide a sufficiently informative uncertainty quantification. \cla{Here, we choose a slightly different way of selecting the weights, where we instead} incorporate a prior belief that, within the given interval of non-zero prior probability, the prior probability of choosing a model with a larger value of $\eta$ is higher, reflecting our prior belief of to what extent parameters drawn from each prior component increases the risk of overfitting in the predictive model. To model this prior distribution, the unnormalised weights in the mixture prior are set to the inverse of the corresponding marginal likelihood\cla{:} 
 \begin{equation}\label{eq:weight_c}
    \omega_c =\frac{{p_c(\mathbf{y})}^{-1}}{\sum_{c=1}^C {p_c(\mathbf{y})}^{-1}}.
 \end{equation}
 
 We are also interested in incorporating prior uncertainty over the exponent of the angular degree $l$ in the prior variance term. To do this, we use a mixture prior with two components. These components, $p_1(\mathbf{z})$ and $p_2(\mathbf{z})$, are both Gaussian, with a specific set of hyperparameters. Specifically, $p_1(\mathbf{z}) = \mathcal{N}(\mathbf{0}, \mathbf{\Omega}^{-1}_1)$ and $p_2(\mathbf{z}) = \mathcal{N}(\mathbf{0}, \mathbf{{\Omega}^{-1}_2})$. Each covariance matrix, $\mathbf{{\Omega}^{-1}_1}$ and $\mathbf{\Omega}^{-1}_2$, is diagonal with parameters \cla{${\Omega_{1}}_{i,i} = \eta_1 l_i$} and ${\Omega_{2}}_{i,i} = \eta_2 l_i^2$, respectively. We fix the prior variance of each component using $\eta_1 = 275.47$ and $\eta_2 = 100.00$, respectively. \cla{These values are chosen using the empirical approach discussed previously, based on independent fits using two separate models with priors adopting each set of hyperparameters, respectively. They provide comparable fits to the data in terms of mean deviation from observed values at the mean of the component-wise posterior distributions.} We use the weights $\omega_1 = \omega_2 = 0.5$ in the mixture prior.

\section{Results}\label{results}
 
 In this section, we present the magnetic field distributions and corresponding uncertainty quantifications obtained using a) fixed prior hyperparameters chosen based on the empirical approach commonly used in standard ZDI, and b) the two probabilistic models extended by a mixture prior, as described in the previous section. \cla{For code and additional implementation details, we refer the reader to Appendix \ref{sec:implementation_details}.}

\subsection{Fixed hyperparameters}

 \begin{figure}[t]
    \centering
    \includegraphics[width=0.82\columnwidth]{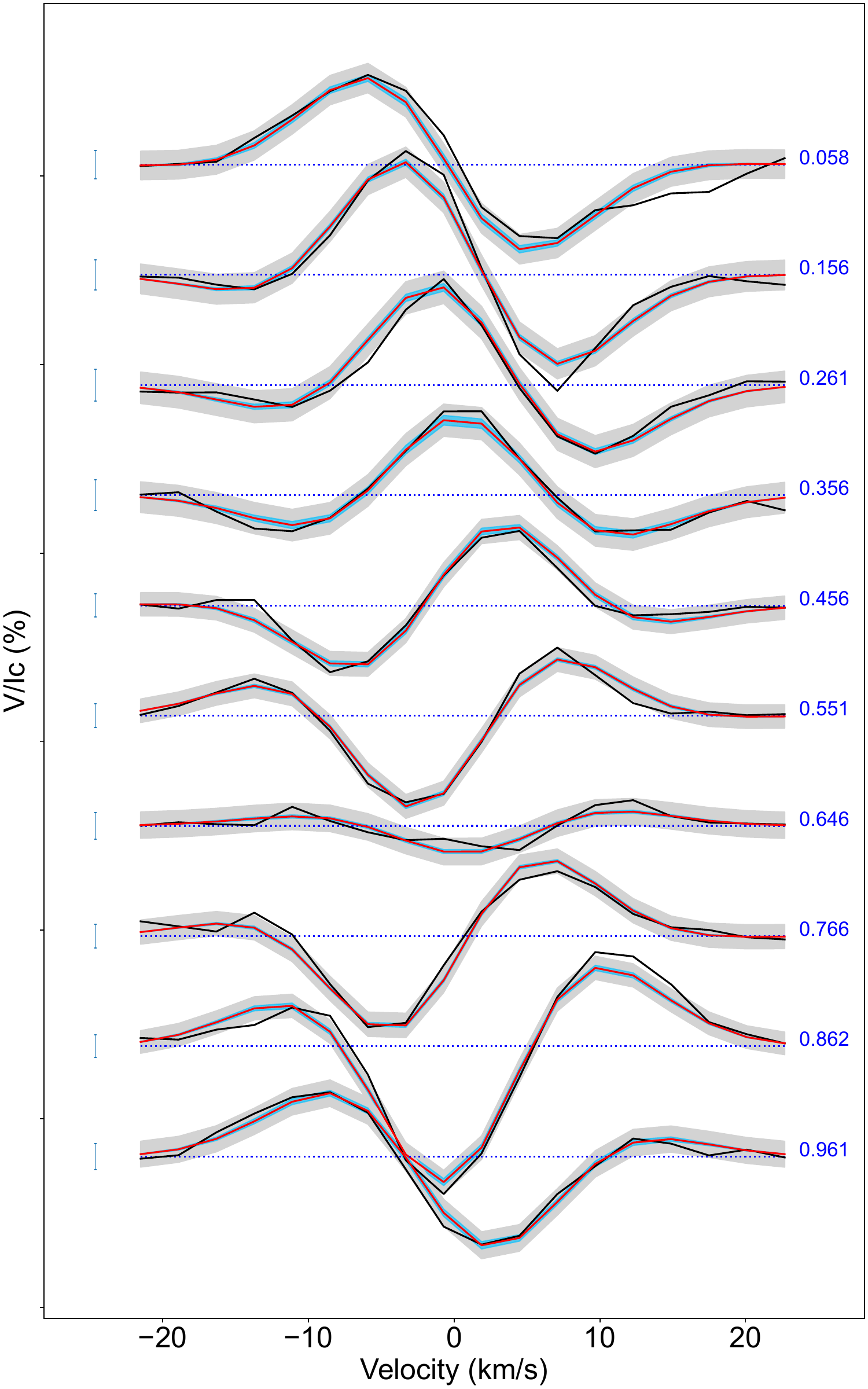}
    \caption{Comparison between the mean LSD Stokes $V$ profiles (red) of the predictive distribution and the observed LSD Stokes $V$ profiles (black), \revb{for a subset of the rotational phases}. The profiles are offset vertically according to the rotational phase, as indicated in blue. \cla{The marginal predictive uncertainty is shaded in light grey and depicts three standard deviations. The corresponding posterior predictive uncertainty is shaded in light blue, and the blue bars represent the corresponding observational uncertainties.}}
    \label{fig:predictive_distribution}
\end{figure}

 This section presents the posterior magnetic field distribution obtained using the statistical model described in Sect.~\ref{model_formulation} with fixed prior hyperparameters. Specifically, $\eta$ is determined according to Sect.~\ref{sec:ZDI}. We use the prior parameters $\Omega_{i,i} = \eta l_i^2$ with $\eta=100$. At the mean of the posterior distribution, this value of $\eta$ yields a quotient of $2.9$ between the \cla{optimised} weighted LS objective value and the regularisation objective value, evaluated according to the objective function in Eq.~\eqref{eq:opt}. A comparison between the observed and modelled LSD Stokes $V$ profiles is illustrated in Fig.~\ref{fig:predictive_distribution}, which shows the best fit at the mean of the predictive distribution\cla{, along with the predictive uncertainty in terms of the marginal standard deviation of the predictive distribution. These quantities are derived from the} predictive distribution given by Eq.~\eqref{eq:pred_dist}, with closed form expression\cla{s for the mean vector and covariance matrix} as detailed in Sect.~\ref{sec:post_inf}. Note that the predictive uncertainty arises from two sources: the posterior uncertainty in the latent variables and the observational noise. \revb{As expected,} the contribution from the posterior \cla{predictive} uncertainty \cla{is \revb{generally} larger} towards the centre of the Stokes $V$ profile at each rotational phase. \revb{See Appendix \ref{sec:app_predictive_distribution} for the corresponding results for all rotational phases.}
 The mean \cla{marginal} standard deviation, derived from the predictive distribution, is $4.98\cdot10^{-5}$ and the fit quality, in terms of mean deviation between the best fit and the observed Stokes $V$ profiles at the mean of the predictive distribution, is $9.17\cdot10^{-5}$. \revb{These results account for all rotational phases.}

\begin{figure}[t]
    \centering
    \includegraphics[width=1\columnwidth]{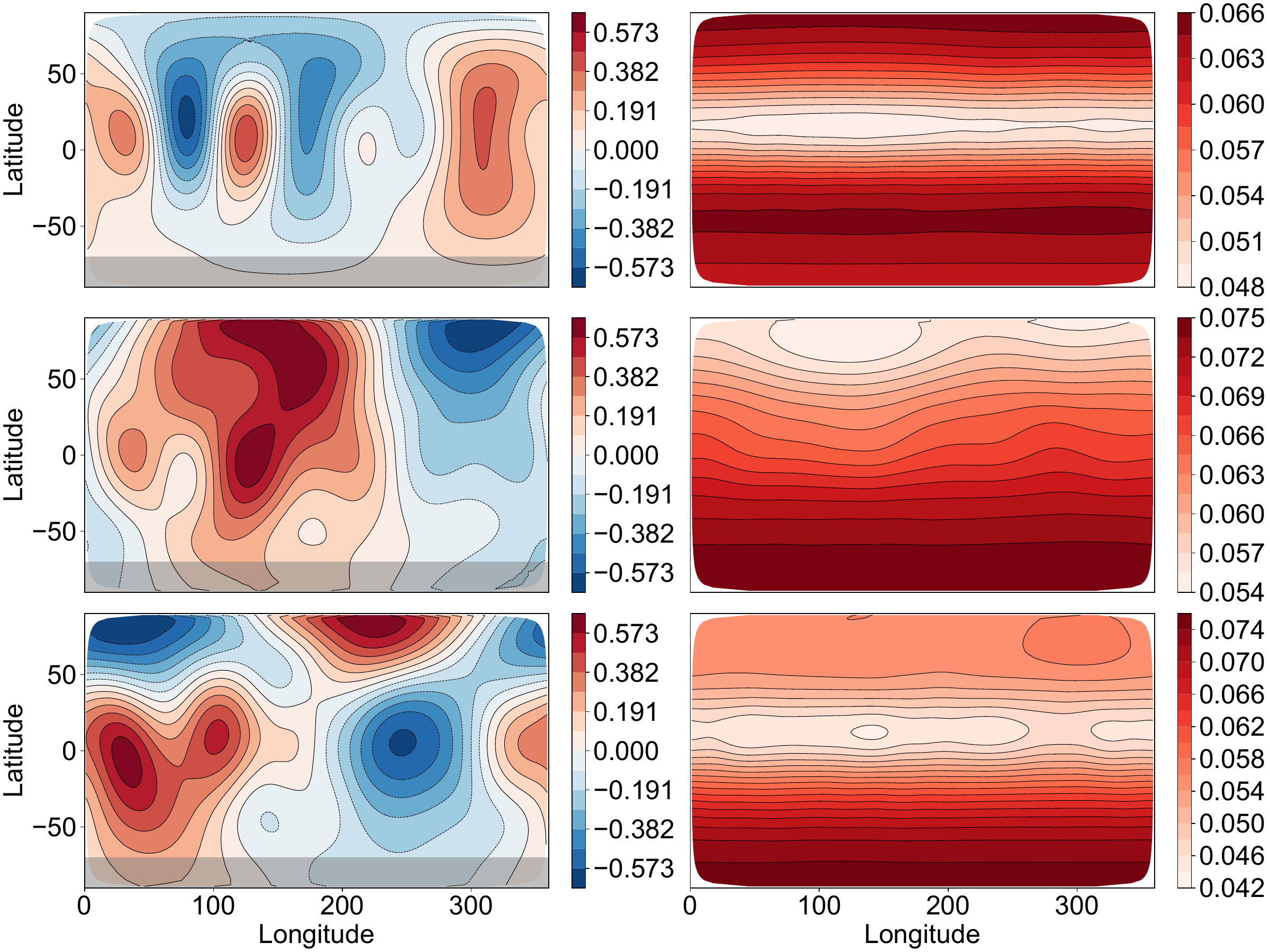}
    \caption{Rectangular maps of the reconstructed magnetic field and the corresponding uncertainty quantification. The left column shows the mean of the posterior magnetic field distribution across the stellar surface, in terms of the radial (top), meridional (middle) and azimuthal (bottom) magnetic field components. The field strength is given in kG. 
    The grey transparent rectangle indicates the  part of the stellar surface that is obscured from view.
    The right column shows the corresponding standard deviation maps. The results are obtained using a statistical model with hyperparameters $\Omega_{i,i} = \eta l_i^2$ and $\eta=100$ in the marginal prior distribution for each spherical-harmonic coefficient $z_i$. Note that the contour plots in this paper consistently use the same number of levels between the minimum and maximum values in all figures.}
    \label{fig:img_l2}
\end{figure}

The \cla{maps of the} mean and standard deviation of the posterior magnetic field distribution are displayed in Fig.~\ref{fig:img_l2}. As evident from these \cla{images}, the surface distribution of the ZDI uncertainties exhibits a distinct latitudinal pattern, reflecting the different sensitivity of the disc-integrated Stokes $V$ profiles to the three components of the magnetic field vector. Specifically, the radial and azimuthal fields are determined most precisely around the sub-observer latitude whereas the meridional field has the least error at the visible rotational pole. At the same time, the precision of the azimuthal field drops off towards the visible pole less sharply than for the radial field. The lack of longitudinal variation of the standard deviation maps is due to a dense rotational phase sampling of the particular observational data set analysed in our paper. In other situations, for example when large phase gaps exist in the data, we would expect our method to yield a noticeable longitudinal variation of uncertainties corresponding to this uneven phase coverage.

In terms of the average standard deviation, we find values in the 0.056--0.067~kG range. This corresponds to 10.4--14.0\% of the peak (95 percentile) values of the respective mean magnetic maps. If we instead consider maximum errors (95 percentile for the visible part of the surface), we obtain 0.065--0.074~kG corresponding to 12.9--15.7\% of the magnetic field amplitudes. These two fractional error estimates illustrate the average (most representative) and maximum (conservative) errors of the ZDI inversion for \sco\ with fixed hyperparameters.

\subsection{Mixture priors}

We analyse the resulting posterior magnetic field distributions obtained using two different mixture priors, capturing prior uncertainty over the exponent of the angular degree $l$ and $\eta$, respectively, in the parameterisation of the prior variance. 

\begin{figure}[t]
    \centering
\includegraphics[width=0.82\columnwidth]{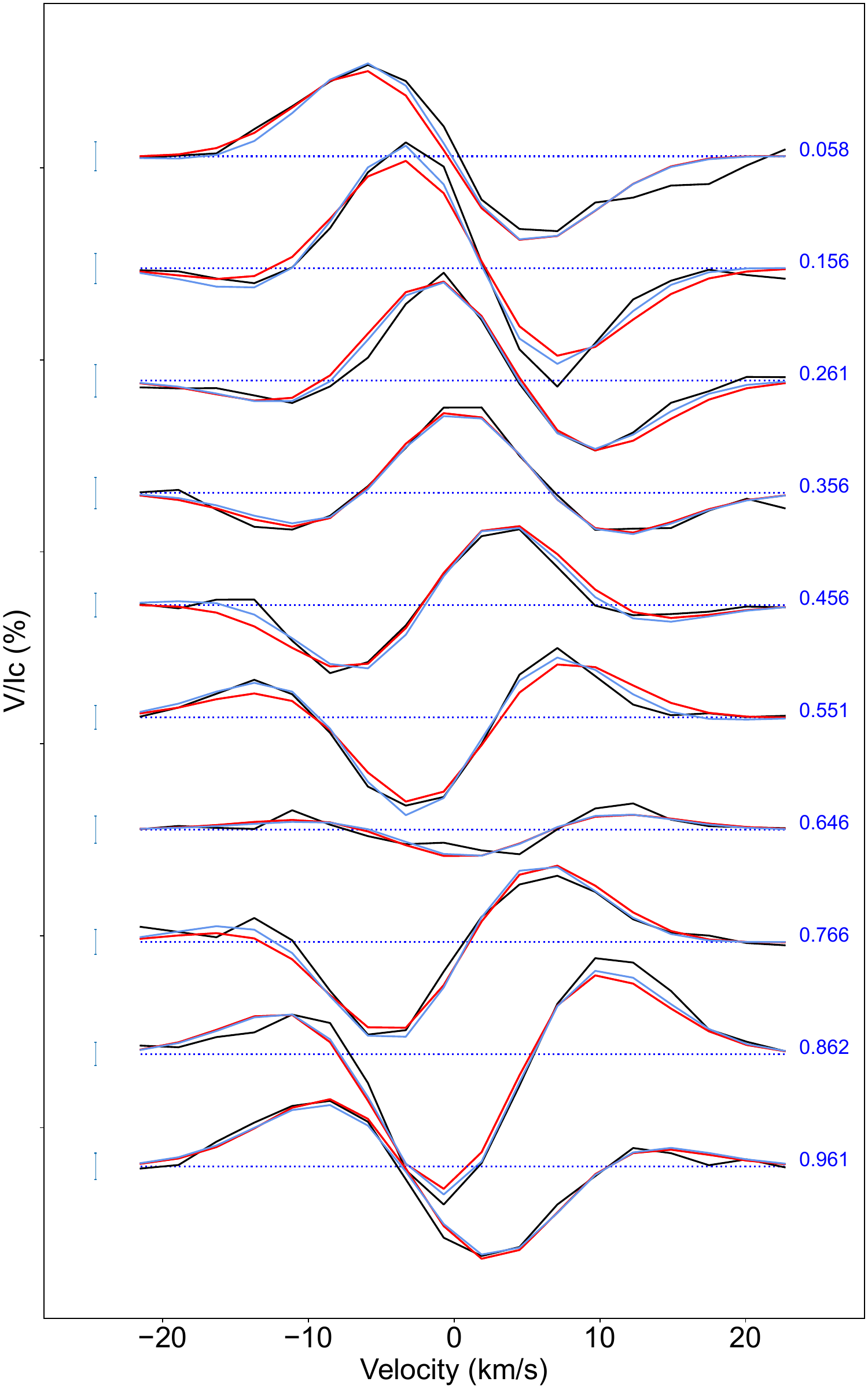}
    \caption{Comparison of predicted (mean) LSD Stokes $V$ profiles from independent inversions using $\eta=16$ (blue) and $\eta=421$ (red) \cla{for a subset of the rotational phases.} The mean deviations between the means of the predictive distributions and the observed Stokes $V$ profiles (black) is $7.87\cdot10^{-5}$ and $1.05\cdot10^{-4}$, respectively, \cla{across all rotational phases}.}
    \label{fig:different_regs}
\end{figure}

First, we present the posterior magnetic field distribution using a mixture prior with 1000 components capturing the prior uncertainty over $\eta$. According to \revb{Sect.~\ref{sec:mixture_priors}}, each prior component $p_c(\mathbf{z}) = \mathcal{N}(\mathbf{0}, \mathbf{{\Omega}^{-1}_c})$ has $\eta$-dependent parameters ${\Omega_{c}}_{i,i} = \eta_c l_i^2$, with parameters $\eta_c$ evenly distributed on the interval $[16,421]$. To illustrate the impact of $\eta$ on the predicted LSD Stokes $V$ profiles, the computed means based on independent inversions using $\eta~=~16$ and $\eta~=~421$, respectively, are presented in Fig. \ref{fig:different_regs}. 
Using the prior weights \cla{given in Eq.~\eqref{eq:weight_c}}, we obtain equal posterior weights for all components in the mixture posterior distribution according to Eq.~\eqref{eq:post3}. The final inversion results in a mean deviation of $9.6\cdot10^{-5}$ between the mean of the predictive distribution and the observed LSD Stokes $V$ profiles, similar in magnitude to the fixed hyperparameter case. The \cla{mean marginal} standard deviation derived from the predictive distribution has increased slightly, reaching $5.16\cdot10^{-5}$. \cla{Appendix \ref{sec:app_predictive_distribution} shows a comparison between the observed LSD Stokes $V$ profiles and the prediction at the mean of the predictive distribution, together with the predictive uncertainty.} 

 \begin{figure}[t]
    \centering
    \includegraphics[width=1.0\columnwidth]{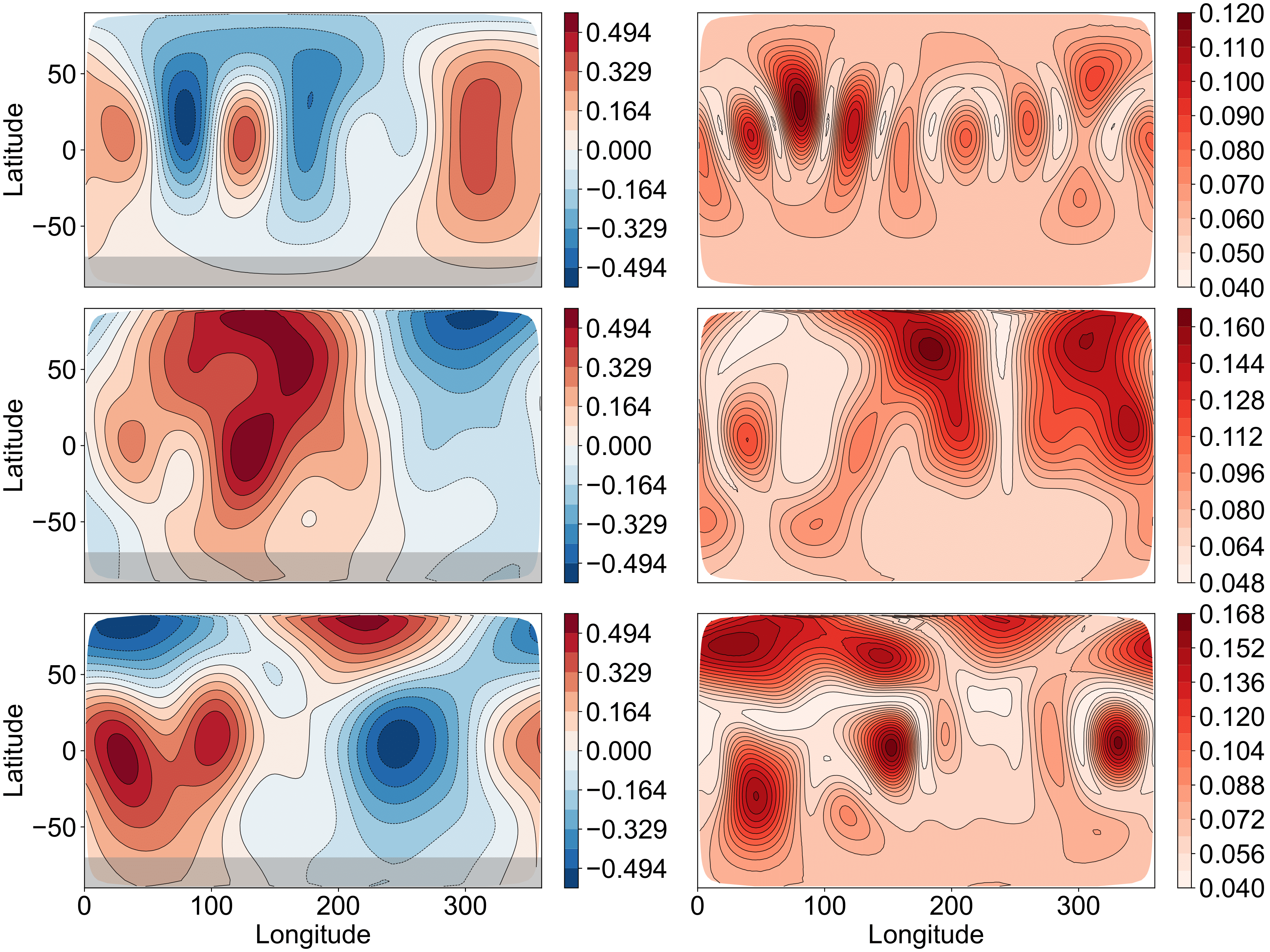}
    \caption{
    Same as Fig.~\ref{fig:img_l2} but for 
    results obtained using a statistical model with a mixture prior consisting of 1000 $\eta$-dependent components $p_c(\mathbf{z})$.}
    \label{fig:img_tau}
\end{figure}

\begin{figure*}[h]
    \centering
    \includegraphics[width=1.0\textwidth]{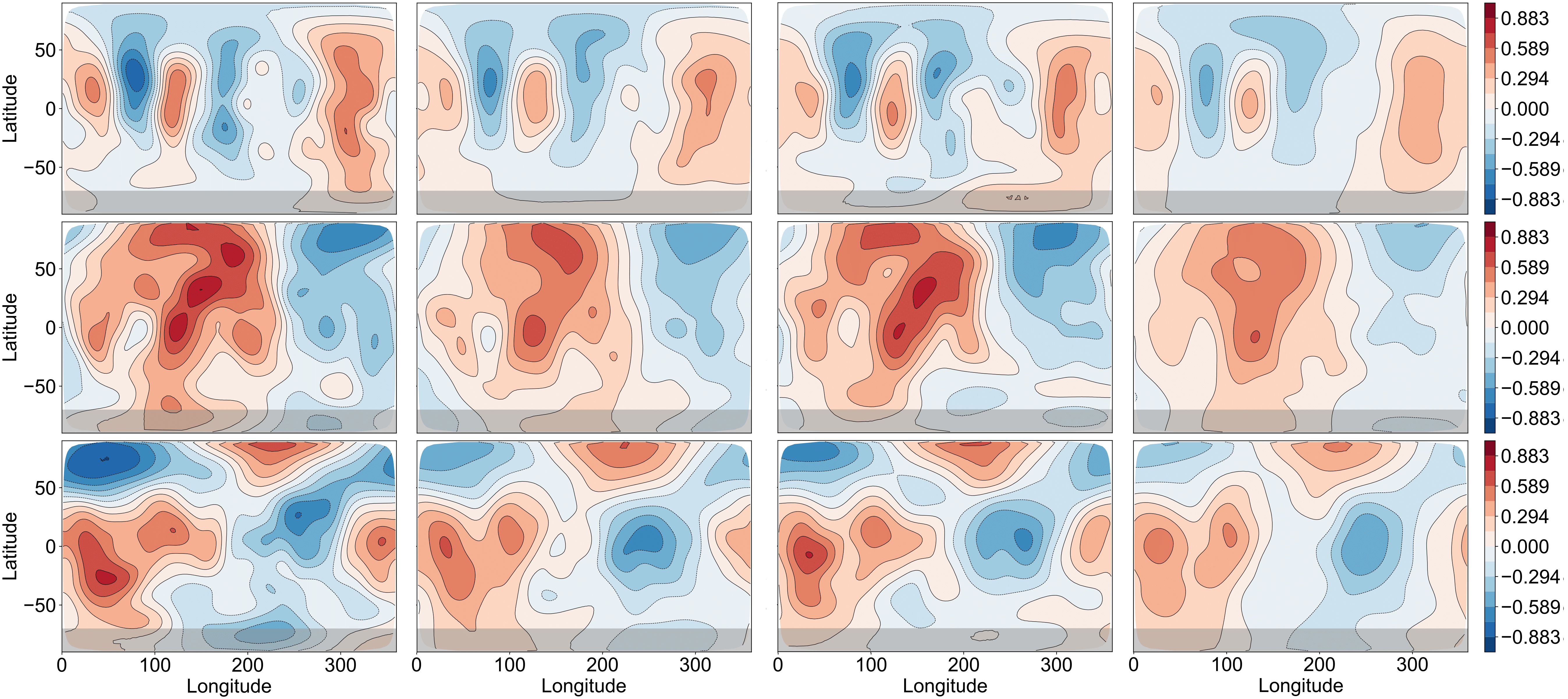}
    \caption{Four samples of surface magnetic field vector maps drawn from the posterior distribution. The distribution is obtained through posterior inference using a statistical model with a mixture prior consisting of 1000 $\eta$-dependent components $p_c(\mathbf{z})$. \cla{From the left, the samples are drawn from components of the mixture posterior distribution with parameters $\eta \approx 34, \eta \approx 123, \eta \approx 62$ and $ \eta \approx  258$, respectively.}
    }
    \label{fig:img_sample_comp}
\end{figure*}

The mean and standard deviation of the posterior magnetic field distribution is calculated as that of the mixture distribution obtained by calculating the posterior magnetic field distribution corresponding to each component in the mixture posterior distribution given in Eq.~\eqref{eq:post3}. Fig. \ref{fig:img_tau} shows the mean of the resulting mixture posterior magnetic field distribution across the stellar surface, together with the corresponding standard deviation maps. Four samples from the resulting distribution are displayed in Fig.~\ref{fig:img_sample_comp}. It can be noted that with $C=1000$ components, the obtained mean magnetic field maps and corresponding uncertainty maps do not change significantly when the number of components increases. Thus, the obtained results closely correspond to using \cla{the corresponding continuous} hyperprior $p(\eta)$ in a hierarchical model.\looseness=-1

As one can see in Fig.~\ref{fig:img_tau}, the surface pattern of the standard deviation changes drastically compared to the results presented in the previous section. Instead of the smooth latitudinal variation seen in Fig.~\ref{fig:img_l2}, we find a highly structured standard deviation map, with the largest scatter roughly corresponding to the strongest (by absolute value) features in the mean magnetic maps. This is to be expected since decreasing $\eta$ generally leads to an increase in the contrast of each map, with sharper details and more prominent small-scale features. Quantifying the standard deviation distribution, we derive the average values 0.063--0.088~kG and maximum values 0.087--0.139~kG. This corresponds to 15.9--18.0\% and 22.2--28.8.0\% of the 95 percentile mean magnetic map amplitudes, respectively -- significantly larger than in the fixed-$\eta$ case.

\begin{figure}[t]
    \centering
    \includegraphics[width=1.0\columnwidth]{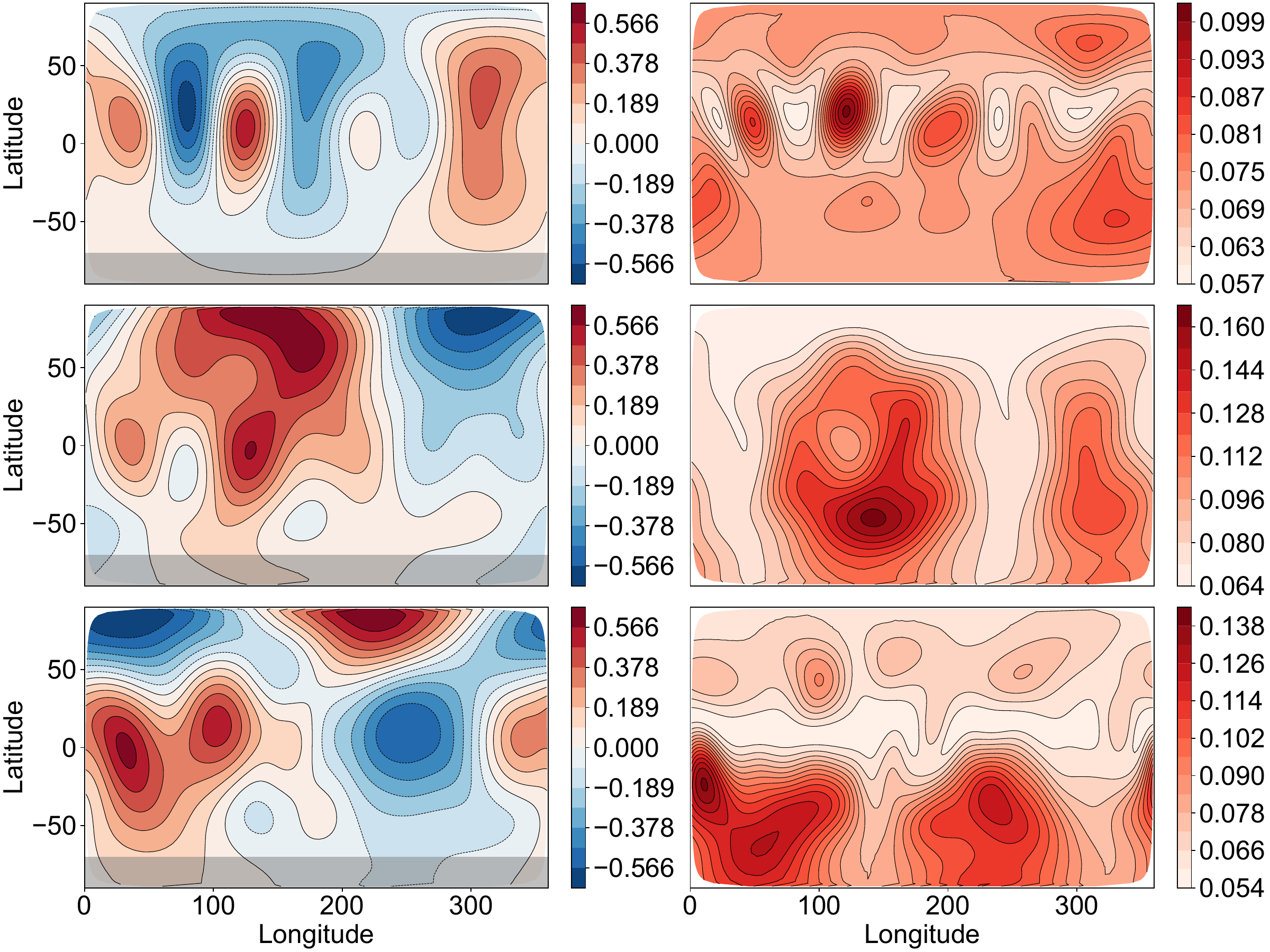}
    \caption{
    Same as Fig.~\ref{fig:img_l2} but for results
    obtained using a statistical model with a mixture prior consisting of the two components 
    $p_1(\mathbf{z})$ and $p_2(\mathbf{z})$.}
    \label{fig:img_l1l2}

\end{figure}

\begin{figure*}[t]
    \centering
\includegraphics[width=1.0\textwidth]{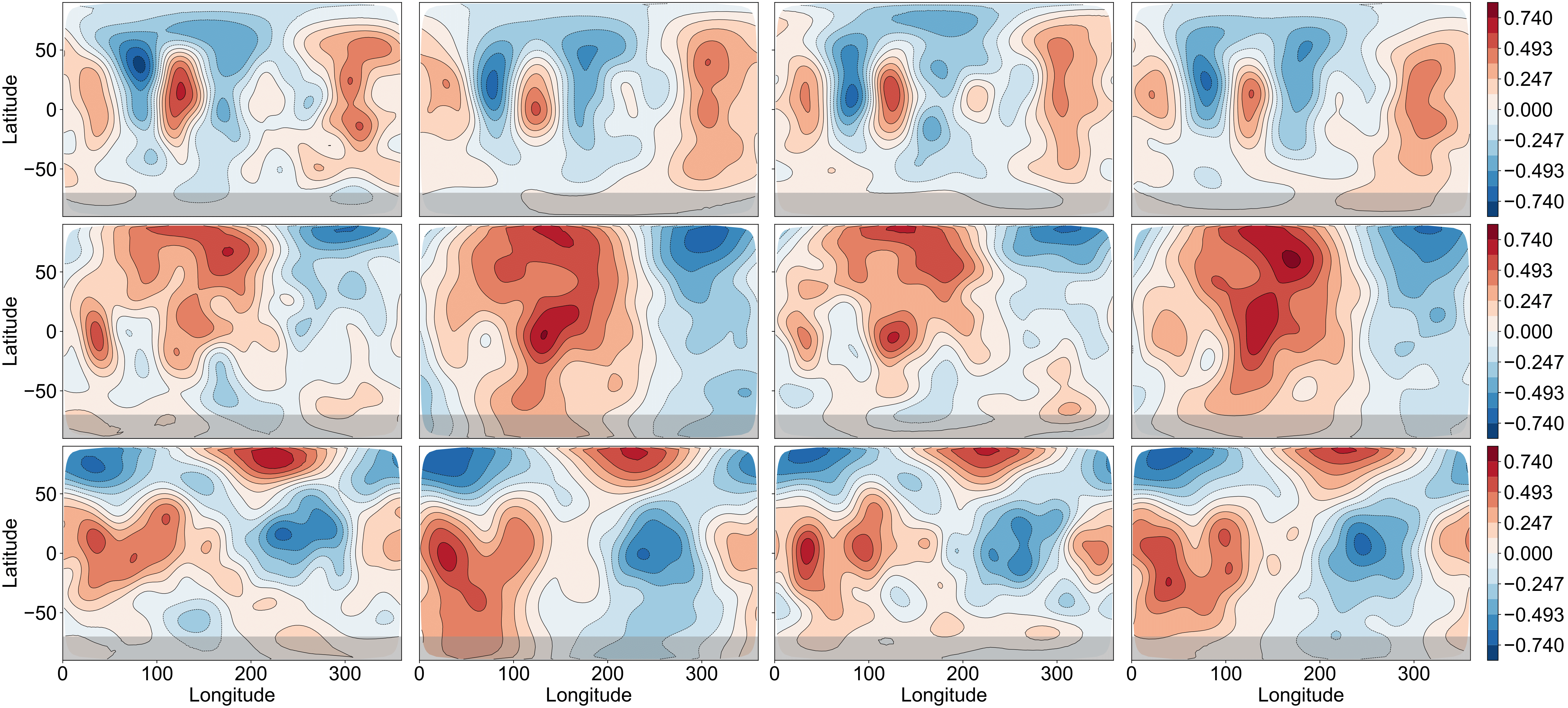}
    \caption{
    Same as Fig.~\ref{fig:img_sample_comp} but the four samples are drawn from the posterior magnetic field distribution obtained through posterior inference using a statistical model with a mixture prior consisting of the two components $p_1(\mathbf{z})$ and $p_2(\mathbf{z})$. \cla{From the left, the first and third column represent samples drawn from the first component of the mixture posterior distribution, and the second and fourth column represent samples drawn from the second component.}
    }
    \label{fig:img_samplel1l2_2}
\end{figure*}

Next, we present the magnetic field distribution corresponding to the two-component mixture prior. Recall that $p_1(\mathbf{z})~=~\mathcal{N}(\mathbf{0}, \mathbf{\Omega}^{-1}_1)$ and $p_2(\mathbf{z}) ~=~\mathcal{N}(\mathbf{0}, \mathbf{{\Omega}^{-1}_2})$, with parameters \cla{${\Omega_{1}}_{i,i} = \eta_1 l_i$} and ${\Omega_{2}}_{i,i} = \eta_2 l_i^2$. With the specific choice of hyperparameters defined in \revb{Sect.~\ref{sec:mixture_priors}}, and prior weights $\omega_1 = \omega_2 = 0.5$, the marginal likelihoods coincide, i.e. $p_1(\mathbf{y}) = p_2(\mathbf{y})$. Consequently, the normalised weights in the resulting mixture posterior becomes ~$\left({\sum_{c=1}^{2} \omega_c p_c(\mathbf{y})}\right)^{-1}\omega_1 p_1(\mathbf{y}) ~= ~\left({\sum_{c=1}^{2} \omega_c p_c(\mathbf{y})}\right)^{-1}\omega_2 p_2(\mathbf{y}) = 0.5$. The mean deviation between the mean of the resulting predictive distribution and the observed LSD Stokes $V$ profiles is $9.09\cdot10^{-5}$. In this case, the mean \cla{marginal} standard deviation of the predictive distribution corresponding to the computed LSD Stokes $V$ profiles is $5.00\cdot10^{-5}$. \cla{See Appendix \ref{sec:app_predictive_distribution} for a comparison between the observed LSD Stokes $V$ profiles and the prediction at the mean of the predictive distribution}.
Fig. \ref{fig:img_l1l2} shows the mean of the posterior magnetic field distribution across the stellar surface, together with the corresponding standard deviation maps. Four samples from the resulting distribution are displayed in Fig.~\ref{fig:img_samplel1l2_2}.  
In this case, the standard deviation maps retain some of the small-scale structures highlighted in the discussion of Fig.~\ref{fig:img_tau}, but these structures are no longer closely associated with the strongest features in the magnetic maps. The representative amplitude of the standard deviation is comparable to the previous test: 0.072--0.098~kG on average and 0.082--0.137~kG if one considers the 95 percentiles, i.e. 15.3--19.5\% and 19.7--27.4\% of the magnetic map amplitudes, respectively. 

\subsection{\cla{Magnetic energy spectrum}}

\begin{figure}[t]
    \centering
\includegraphics[width=\columnwidth]{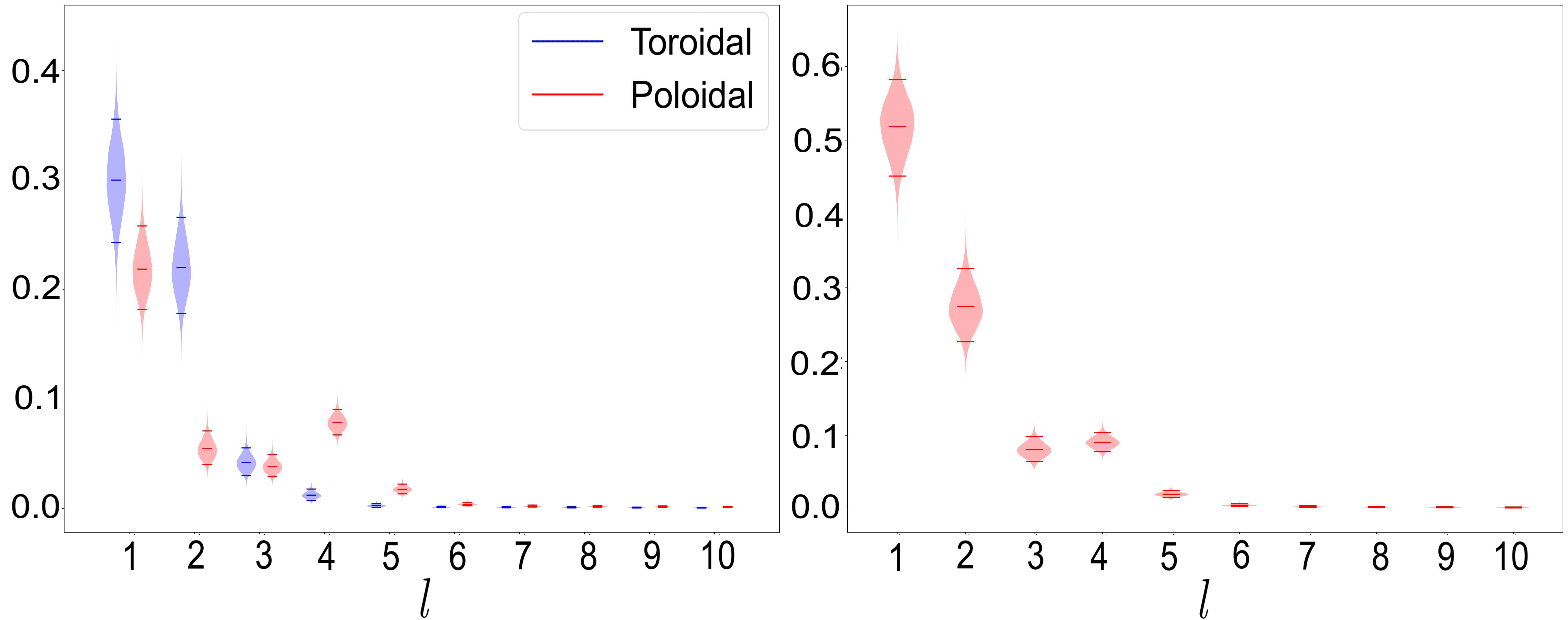}
    \caption{Violin plots illustrating the distribution of magnetic energy over different harmonic modes.
    \cla{In these plots the widths of the coloured regions represent the smoothed probability density of the fraction of the total magnetic energy, numerically obtained by sampling} from the posterior distribution $p(\mathbf{z}|\mathbf{y})$ corresponding to the \cla{results illustrated} in Fig.~\ref{fig:img_l2}. Horizontal lines mark the 0.05-\cla{, 0.50-} and 0.95-quantiles of the distribution at each mode, respectively. The left panel compares the energy of the toroidal contribution (blue) and the poloidal components (red) as a function of angular degree $l$, with values normalised by the total energy, i.e., $(E_{\rm t}, E_{\rm p})/E_{\rm tot}$. The right panel shows the total magnetic energy of the poloidal and toroidal field components as a function of angular degree $l$, also normalised by the total energy, i.e. $(E_{\rm p} + E_{\rm t})/E_{\rm tot}$.}
    \label{fig:SHenergyL2}
\end{figure}

\begin{figure}[t]
    \centering
\includegraphics[width=\columnwidth]{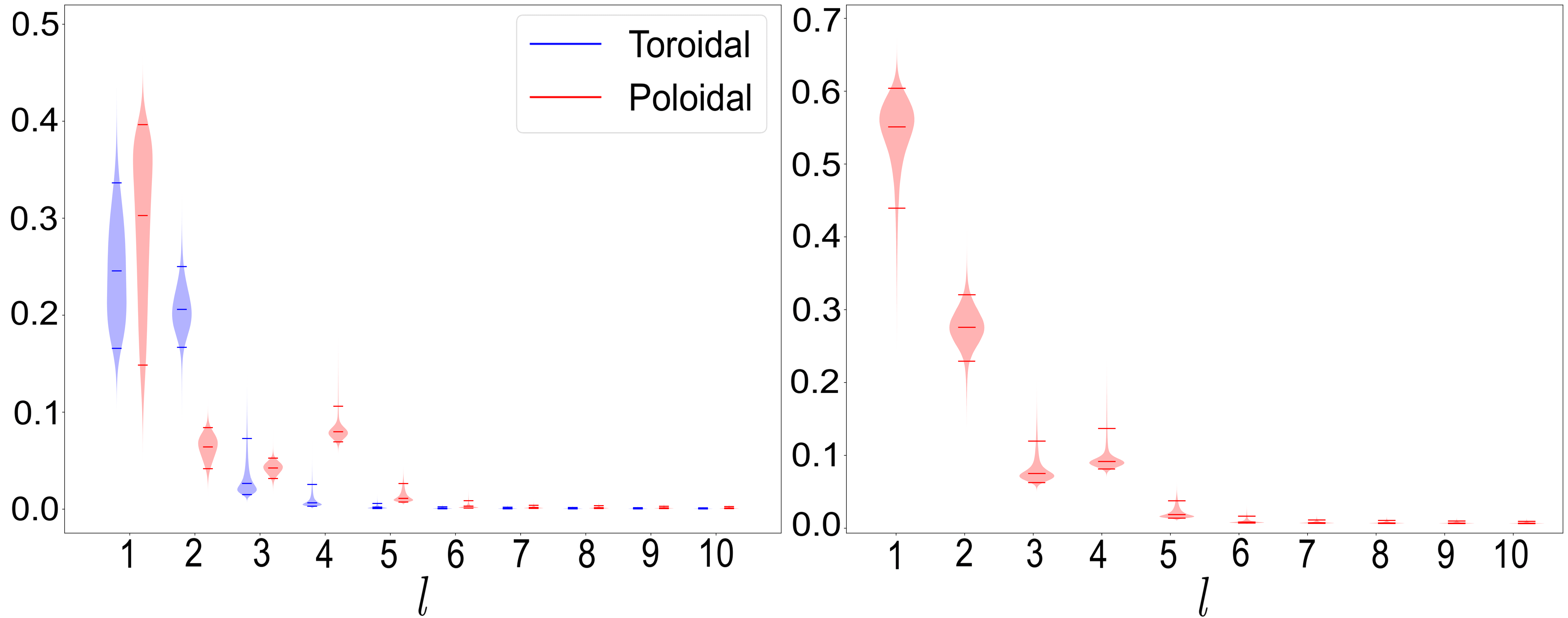}
    \caption{Same as Fig. \ref{fig:SHenergyL2} but for the results obtained from the posterior distribution $p(\mathbf{z}|\mathbf{y})$ corresponding to the magnetic field in Fig.~\ref{fig:img_tau}.}
    \label{fig:SHenergyComp}
\end{figure}

\begin{figure}[t]
    \centering
\includegraphics[width=\columnwidth]{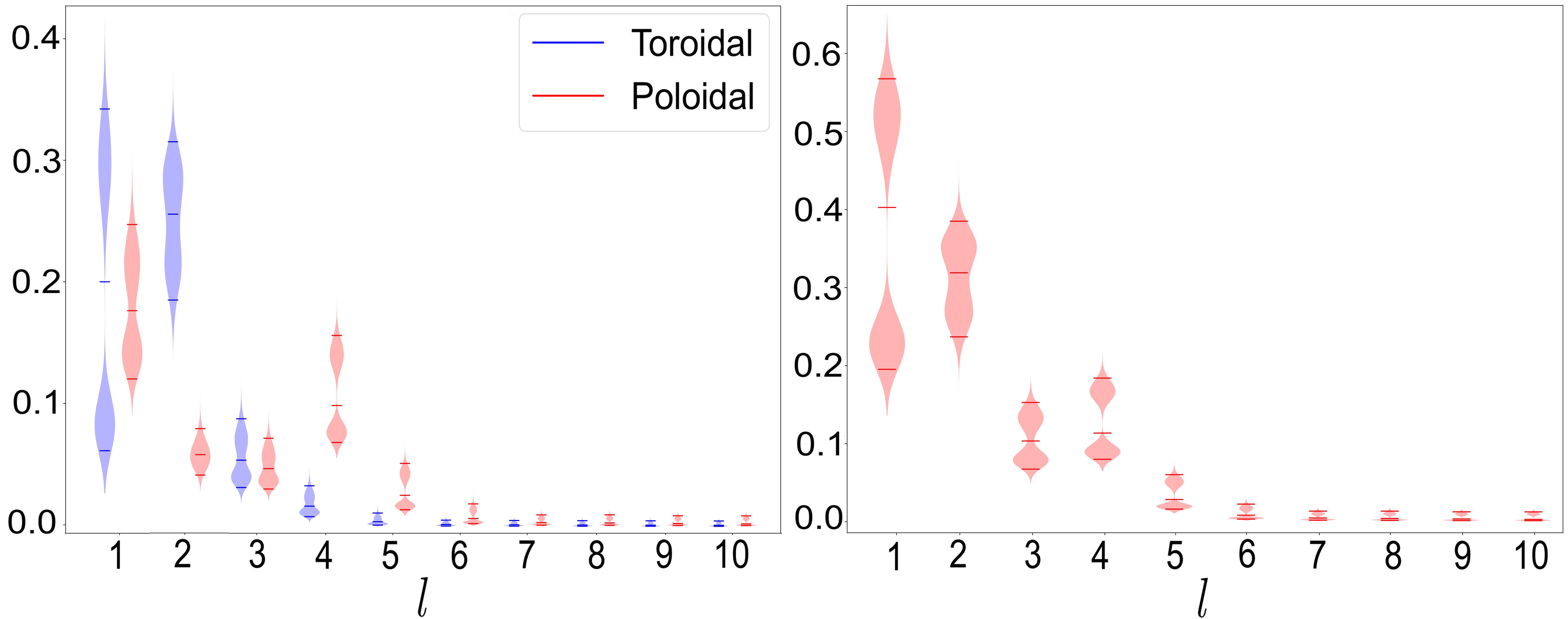}
    \caption{Same as Fig. \ref{fig:SHenergyL2} but for the results obtained from the posterior distribution $p(\mathbf{z}|\mathbf{y})$ corresponding to the magnetic field in Fig.~\ref{fig:img_l1l2}.}
    \label{fig:SHenergyL1L2}
\end{figure}

Using \cla{our} spherical-harmonic field parameterisation, it is straightforward to assess the strength of the axisymmetric vs. non-axisymmetric field, as well as the strength of the poloidal vs. toroidal field components. Specifically, the $\gamma_{l, m}$ coefficients specify the strength of the toroidal field component, whereas the $\alpha_{l, m}$ and $\beta_{l, m}$ coefficients specify the vertical and horizontal components of the poloidal field, respectively.
Given the posterior distribution obtained from each independent choice of prior distribution, we have analysed the \cla{numerical} distribution of the magnetic field energy contributions over different harmonic modes ($l$-modes). These distributions, together with a comparison of the distributions of the magnetic energy of the poloidal and toroidal field components over the harmonic modes, are illustrated in Fig. \ref{fig:SHenergyL2}--\ref{fig:SHenergyL1L2}.
Fig. \ref{fig:SHenergyL2} shows the distributions corresponding to the fixed hyperparameter case, with $\Omega_{i,i} = \eta l_i^2$ and $\eta=100$. These magnetic energies correspond to the posterior magnetic field distribution in Fig. \ref{fig:img_l2}. As we can see, the total magnetic energy is spread over modes $l \in [1,5]$, with a clear peak at $l=1$. Considering the poloidal and toroidal field components separately, we note a significant peak in the poloidal field component at $l=1$, followed by a smaller peak at $l=4$. In contrast, the peak of the toroidal field component ranges over $l \in [1,2]$, whereafter the contribution decreases with $l$.
Fig. \ref{fig:SHenergyComp} shows the distribution of magnetic energy over the harmonic modes corresponding to the posterior magnetic field distribution presented in Fig. \ref{fig:img_tau}, obtained from a mixture prior with with 1000 $\eta$-dependent components $p_c(\mathbf{z})$. 
\cla{Finally,} Fig. \ref{fig:SHenergyL1L2} shows the corresponding results for the posterior magnetic field distribution given in Fig. \ref{fig:img_l1l2}, obtained from a mixture prior with components $p_1(\mathbf{z})$ and $p_2(\mathbf{z})$.

The magnetic energy distributions over harmonic modes illustrated in Fig.~\ref{fig:SHenergyComp}--\ref{fig:SHenergyL1L2} are qualitatively similar to the distributions in the fixed hyperparameter case illustrated in Fig.~\ref{fig:SHenergyL2}. Considering the comparatively large scatter in the resulting magnetic energy based on statistical models with mixture priors, our results indicate high uncertainty in the relative magnetic energies recovered in the ZDI analysis of \sco.

\cla{In addition to analysing the magnetic energy distribution for poloidal vs toroidal field components, we divide the spherical-harmonic modes into two other groups: those with $\vert m \vert < l/2$ and those with $\vert m \vert \geq l/2$ \citep{fares2009}. A statistical summary of the magnetic energy distributions in terms of a) the total contribution from the poloidal field component across all harmonic modes, and b) the isolated contribution from the $|m| < l/2$ field component, is presented in Table \ref{tab:table_magnetic_energy} for each of the three cases.}

\cla{While the magnetic energy, as discussed here, is more interpretable and comparable between studies, we have included additional visualisations of the posterior distribution over $\mathbf{z}$ in Appendix \ref{appendix_c} for completeness.}

\begin{table}[t!]
\caption{Quantiles of the posterior magnetic energy fractions.}
\label{tab:table_magnetic_energy}
\centering
\begin{tabular}{lcccccc}
\hline
\hline
 & \multicolumn{3}{c}{Poloidal field} & \multicolumn{3}{c}{$|m| < l/2$} \\ \cmidrule(lr){2-4} \cmidrule(lr){5-7}
quantile & 0.05 & 0.50 & 0.95 & 0.05 & 0.50 & 0.95 \\
\hline
Case 1   &  0.38  &  0.42  &  0.46  & 0.09 &  0.12 & 0.15 \\
Case 2  &  0.38  & 0.50  & 0.61 &  0.09 & 0.13 & 0.16 \\
Case 3 & 0.39 & 0.46  & 0.53 & 0.09 & 0.15 & 0.19 \\
\hline
\end{tabular}
\end{table}
\tablefoot{\cla{Quantiles of the posterior distribution of the fraction of the total magnetic energy for a) the poloidal field components across harmonic modes and b) the $|m| < l/2$ field component. Case 1 refers to the fixed hyperparameter case, Case 2 refers to the distribution obtained from a mixture prior with 1000 $\eta$-dependent components $p_c(\mathbf{z})$, and Case 3 refers to the distribution obtained from a mixture prior with components $p_1(\mathbf{z})$ and $p_2(\mathbf{z})$}.}

\section{Conclusions and discussion}\label{discussion}
 
In this study, we have presented magnetic field distributions and corresponding uncertainty maps resulting from \cla{a probabilistic} ZDI analysis of the early-type star \sco. We considered three distinct prior distributions over the latent variables in the spherical-harmonic field parameterisation. When the hyperparameters in the proposed prior distribution were chosen using the empirical Bayes approach, our investigation showed that the amplitude in the posterior magnetic field distribution violated the weak-field approximation motivating our choice of forward mapping function $\mathbf{f}(\mathbf{z})$. \cla{We concluded that maximum likelihood estimation of the hyperparameter $\eta$ is ill-posed in this case}. The uncertainty maps obtained with empirically fixed hyperparameters exhibited a smooth latitudinal variation across the stellar surface. By introducing mixture priors, we accounted for prior uncertainty over \cla{two} hyperparameters in the original prior distribution, capturing the sensitivity of the magnetic field distribution to the specific choice of prior and, by extension, to the hyperparameters used in the classical ZDI framework. A mixture prior over $\eta$ increased the structure in the uncertainty maps and raised the overall uncertainty level, particularly around the features in the mean magnetic map with the strongest amplitudes. A similar uncertainty level was observed when introducing a mixture prior over the exponent on the angular degree $l$. Although small-scale structures remained, the correlation with the strongest magnetic features in the mean magnetic field distribution was less prominent.\looseness=-1  

\cla{We also analysed the magnetic energy spectrum, with emphasis on the posterior magnetic energy distribution across $l-$modes.} Compared to previous ZDI inversions targeting \sco, a star \cla{assumed} to exhibit a complex, non-dipolar surface magnetic field, the fact that the magnetic energy contribution is dominated by $l=1$ is unexpected. This result alone would indicate a \cla{predominantly} dipolar field geometry, in line with the majority of hot, magnetic stars, \cla{but in contrast to} previous studies of \sco. \cla{We investigated this issue by an in-depth comparison between our results and the magnetic field map obtained by \citet{kochukhov:2016} with the same harmonic parameterisation (``model 2'' in that paper). It turns out that the overall morphology of the surface distributions of the three vector magnetic components is qualitatively similar in the two studies. However, their relative strengths differ. Here we recover about 20--30\% less relative energy in the radial and azimuthal components, which are dominated by higher $l$ modes. Conversely, our meridional field contributes more than twice the relative energy to the total field compared to the results by \citet{kochukhov:2016}. This meridional field component features a simpler, dipolar-like geometry (see Fig.~\ref{fig:img_l2}). Consequently, the contribution of the $l=1$ harmonics is significantly higher in our results. This modification of the outcome of the ZDI analysis is sufficient to noticeably alter the resulting harmonic energy spectrum. While \citet{kochukhov:2016} found the poloidal field energy to be spread over $l \in [1,4]$ and observed a peak in the toroidal field contribution at $l \in [2,3]$, we recover a stronger toroidal component peaking at $l=1$ with a less important poloidal field peaking at both $l=1$ and $l=4$.}

\cla{Since there are differences in modelling choices between the ZDI reconstructions, we do not expect to exactly reproduce the results at the mean of the posterior magnetic field distribution in this case. In general,} it is worth remarking that the uncertainty quantification obtained from the proposed framework for probabilistic ZDI is conditioned on our choice of probabilistic model, i.e. $p(\mathbf{y}|\mathbf{z})$ and $p(\mathbf{z})$, and the information content in the observations $\mathbf{y}$. As noted in Sect.~\ref{sec:introduction}, earlier studies of the surface magnetic field of \sco\ have compared the field geometries obtained using different spherical-harmonic magnetic field parameterisations, ultimately concluding that the topological details in the surface magnetic field, as well as the average magnetic field strengths, vary significantly depending on the parameterisation \citep{kochukhov:2016}. Since the Bayesian framework naturally extends to Bayesian model selection and Bayesian model averaging, \cla{an} interesting avenue for future work is to perform probabilistic ZDI analyses of \sco\ using competing field parameterisations to model the likelihood.

While this study has focused on ZDI inversion in the weak-field limit, \cla{another} important direction for future research is to investigate the feasibility of employing probabilistic ZDI to generate reliable uncertainty maps when the line-profile response cannot be modelled under this assumption. Addressing this challenge will require advanced computational techniques, such as MCMC methods or variational Bayes \citep{variational_inference_review}, since the posterior distribution $p(\mathbf{z}|\mathbf{y})$ cannot be expressed in closed form for non-linear response profiles under standard model assumptions. Additionally, the proposed framework for probabilistic ZDI can be expanded by extending the hierarchy in the statistical model (see Sect. \ref{hyperpriors_section}) to account for prior uncertainty in stellar parameters beyond the spherical-harmonic coefficients. Such an extension would enable quantification of uncertainty in the surface magnetic field of more complex ZDI targets, including stars in eclipsing binary systems and equator-on hosts of transiting exoplanets, where degeneracy in the latent variables within the spherical-harmonic formulation is likely to occur. Both of these avenues will be pursued in future work.

\cla{To summarise our study, w}e have proposed a Bayesian framework for probabilistic ZDI allowing for formal uncertainty quantification of the obtained stellar magnetic field maps, given a single set of spectropolarimetric observations. Coupling the magnetic field distributions with interpretable uncertainty maps makes it possible to reason about the uncertainty in the obtained magnetic field distributions in a meaningful way. This is something that has not previously been possible, as classical ZDI has generally been restricted to point estimates. We have demonstrated that, for stars exhibiting relatively weak surface magnetic fields such that the weak-field approximation can be used to model the line-profile response, a closed-form solution for the posterior distribution over the spherical-harmonic coefficients exists under specific model assumptions. Our choice of probabilistic model is convenient in the sense that it makes it easy to quickly conduct Bayesian analyses of the surface magnetic fields of a large group of stars, despite the high-dimensional spherical-harmonic magnetic field parameterisation. Since we obtain an uncertainty quantification centered around the point estimate obtained using standard ZDI, our probabilistic ZDI formulation essentially provides uncertainty quantification as a byproduct compared to standard ZDI, without loss of information. 

\begin{acknowledgements}
 This work was supported by the Wallenberg AI, Autonomous Systems and Software Program (WASP) funded by the Knut and Alice Wallenberg Foundation. Computations were enabled by the Berzelius resource provided by the Knut and Alice Wallenberg Foundation at the National Supercomputer Centre. O.K. acknowledges support by the Swedish Research Council (grant agreements no. 2019-03548 and 2023-03667), the Swedish National Space Agency, and a sabbatical fellowship from AI4Research at Uppsala University.
\end{acknowledgements}

\bibliographystyle{aa}
\bibliography{references}

\begin{appendix}
\onecolumn
\FloatBarrier
\section{\cla{Implementation details}}\label{sec:implementation_details}

\cla{Our probabilistic ZDI framework is implemented in Python, utilizing NumPy for general numerical computations and JAX for efficient linear algebra and automatic differentiation. In particular, JAX's automatic differentiation functionality is employed to compute the design matrix $A$ in the forward model, a key component in the Bayesian formulation.  For further implementation details, see our probabilistic ZDI code made available on GitHub\footnote{\url{https://github.com/jenan007/ProbabilisticZDI/}}.}

\FloatBarrier
\section{\cla{Predictive distribution}}\label{sec:app_predictive_distribution}

\begin{figure}[h!]
    \centering
    \includegraphics[width=0.49\columnwidth]{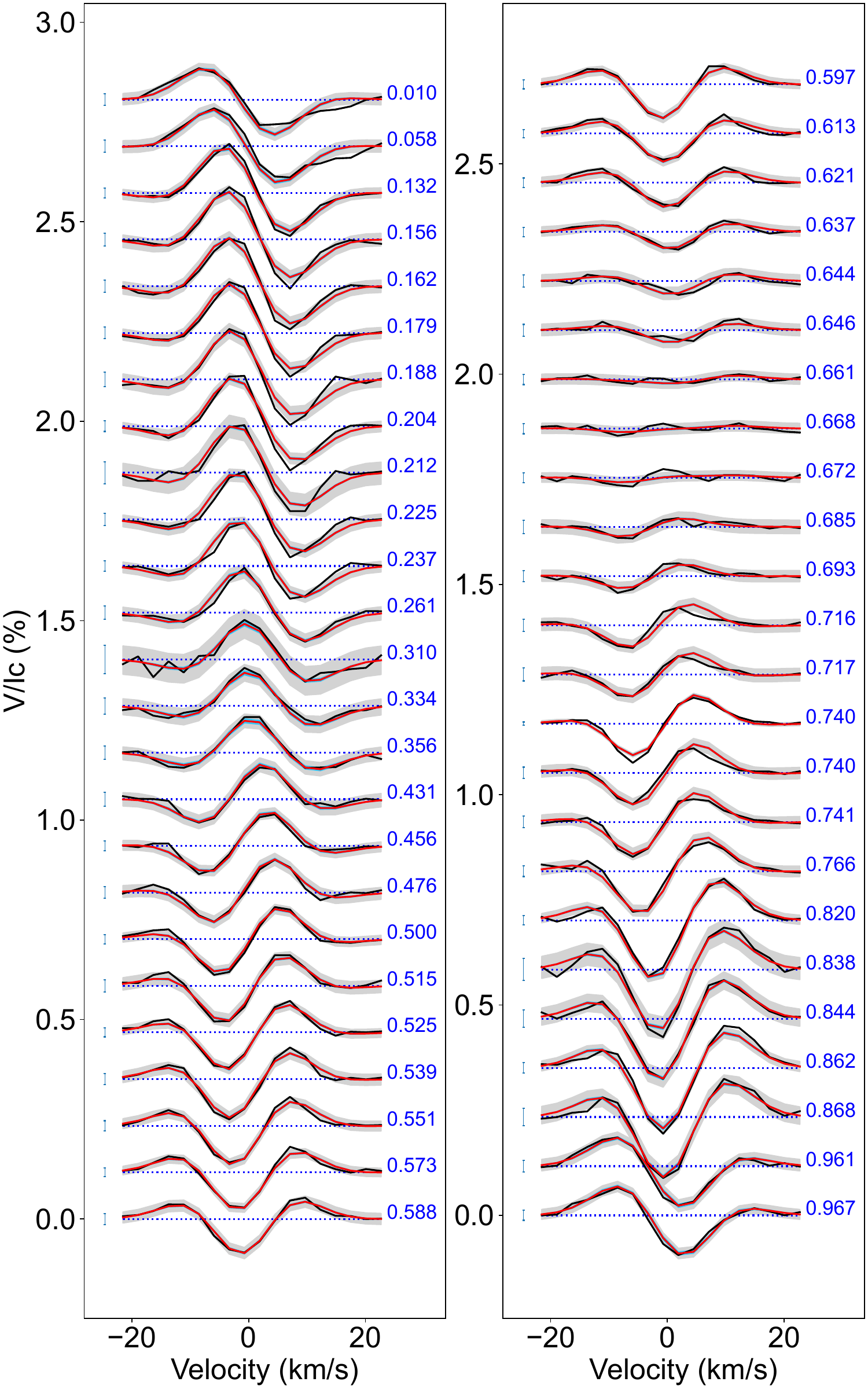}
    \caption{\cla{Same as Fig. \ref{fig:predictive_distribution} but for all rotational phases. \revb{Recall that the observed profiles are depicted in black, and the reconstruction at the mean is depicted in red. T}he shaded light grey area shows the marginal predictive uncertainty, whereas the shaded light blue area shows the corresponding posterior
predictive uncertainty.}}
    \label{fig:predictive_distribution_subset}
\end{figure}

\cla{In Fig. \ref{fig:predictive_distribution}, we showed a comparison between the observed LSD Stokes $V$ profiles and the reconstruction at the mean of the predictive distribution \revb{for a subset of the rotational phases}, together with the posterior and predictive uncertainties. Fig. \ref{fig:predictive_distribution_subset} shows the corresponding results \revb{for all rotational phases}.}

\begin{figure*}[h!]
\centering
\includegraphics[width=0.49\textwidth]{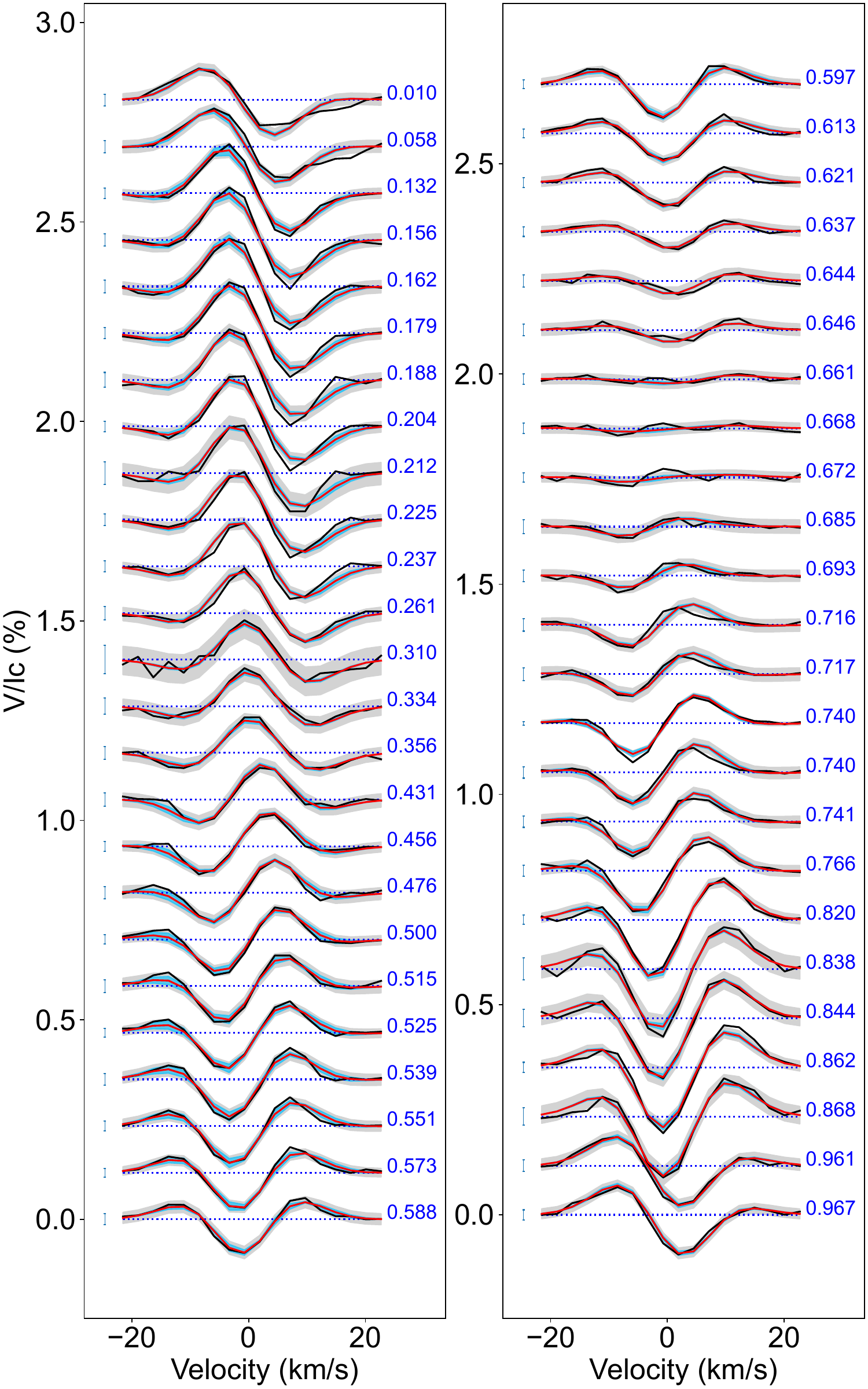}
\includegraphics[width=0.49\textwidth]{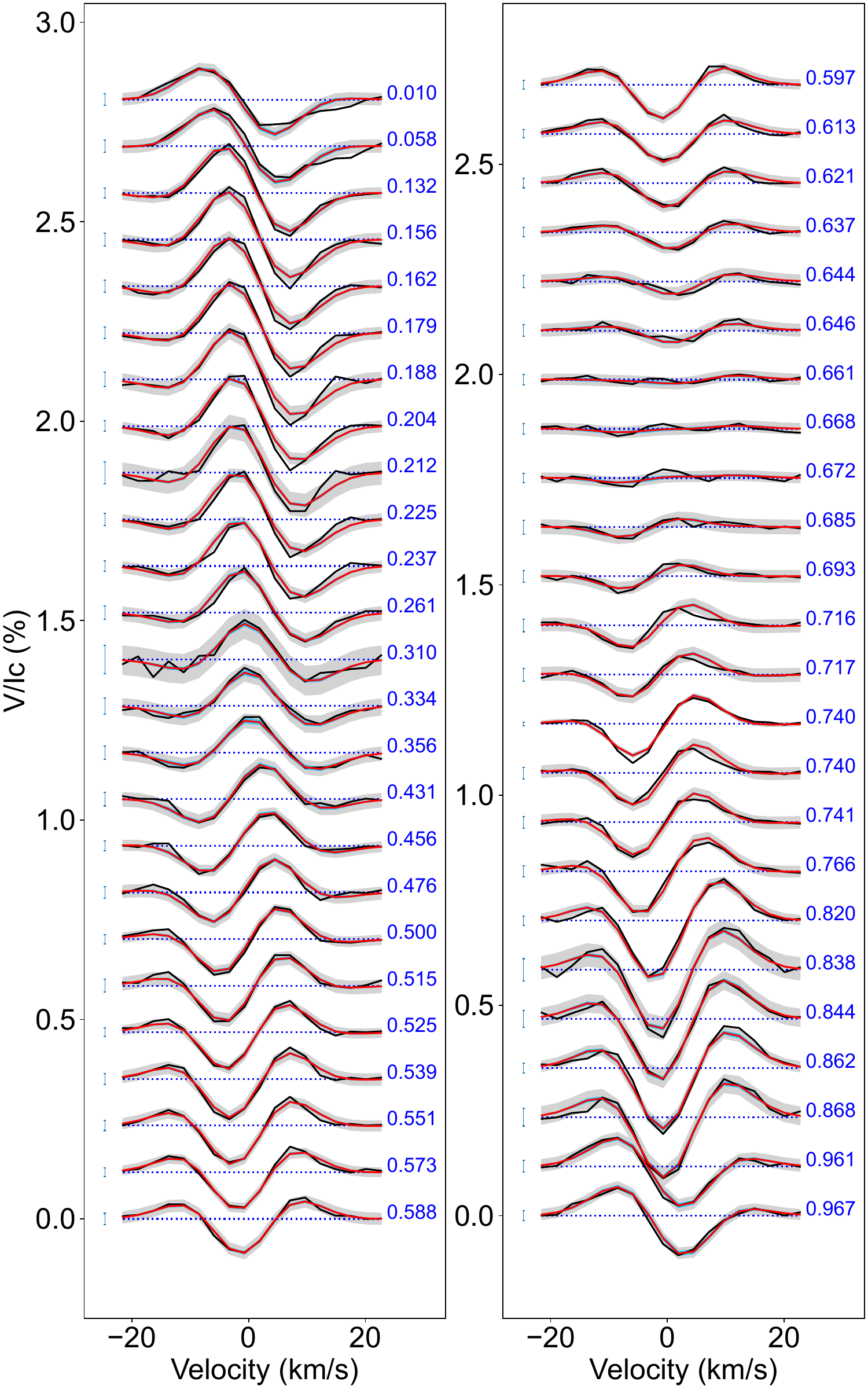}
\caption{\cla{Same as Fig.~\ref{fig:predictive_distribution_subset} but for results obtained using a statistical model with \textbf{a)} a mixture prior consisting of 1000 $\eta$-dependent components $p_c(\mathbf{z})$. and \textbf{b)} a mixture prior consisting of the two components $p_1(\mathbf{z})$ and $p_2(\mathbf{z})$.}}
\label{fig:predictive_distribution L1L2_comp}
\end{figure*}

\begin{figure*}[h!]
\centering
\includegraphics[width=0.49\textwidth]{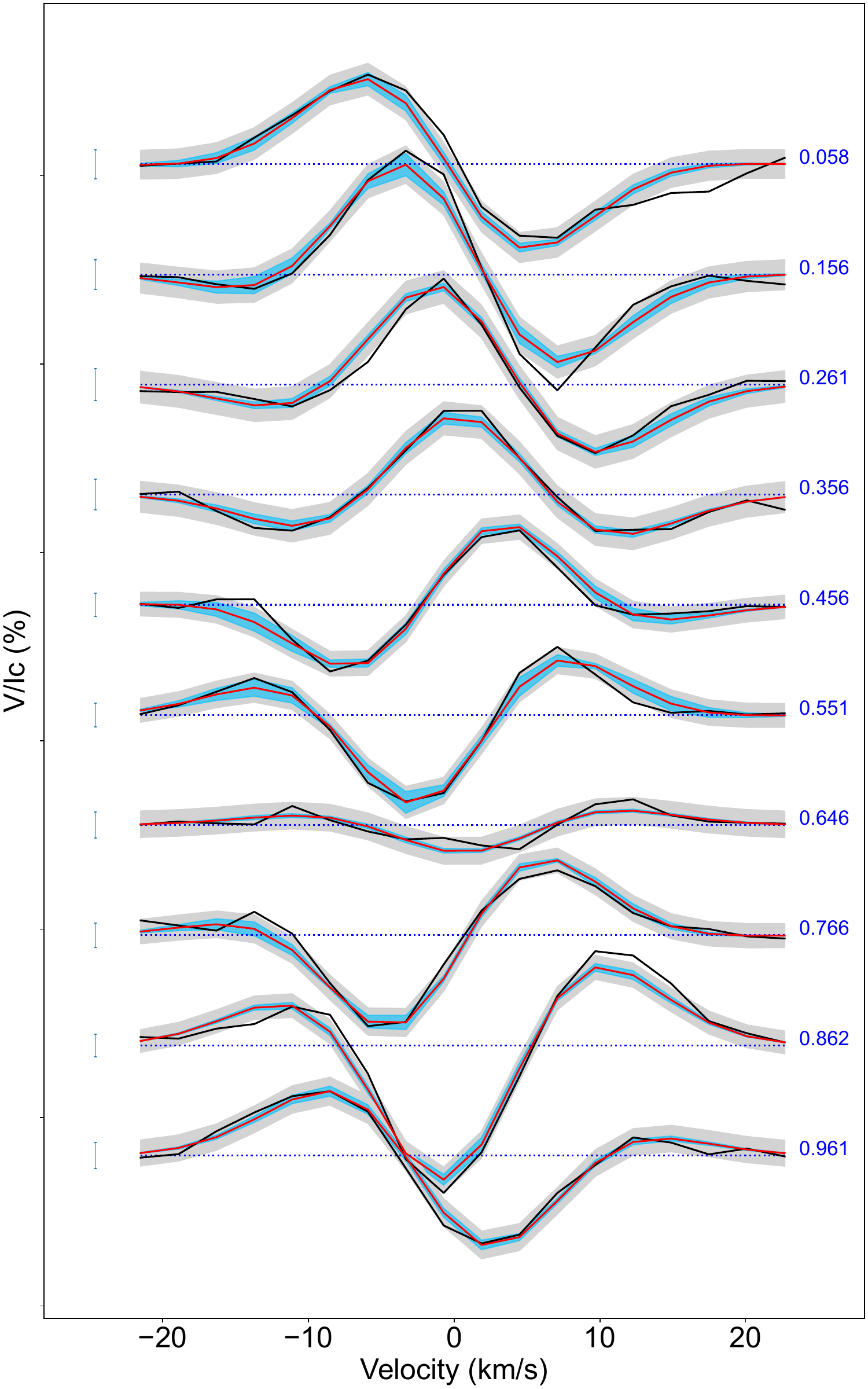}
\includegraphics[width=0.49\textwidth]{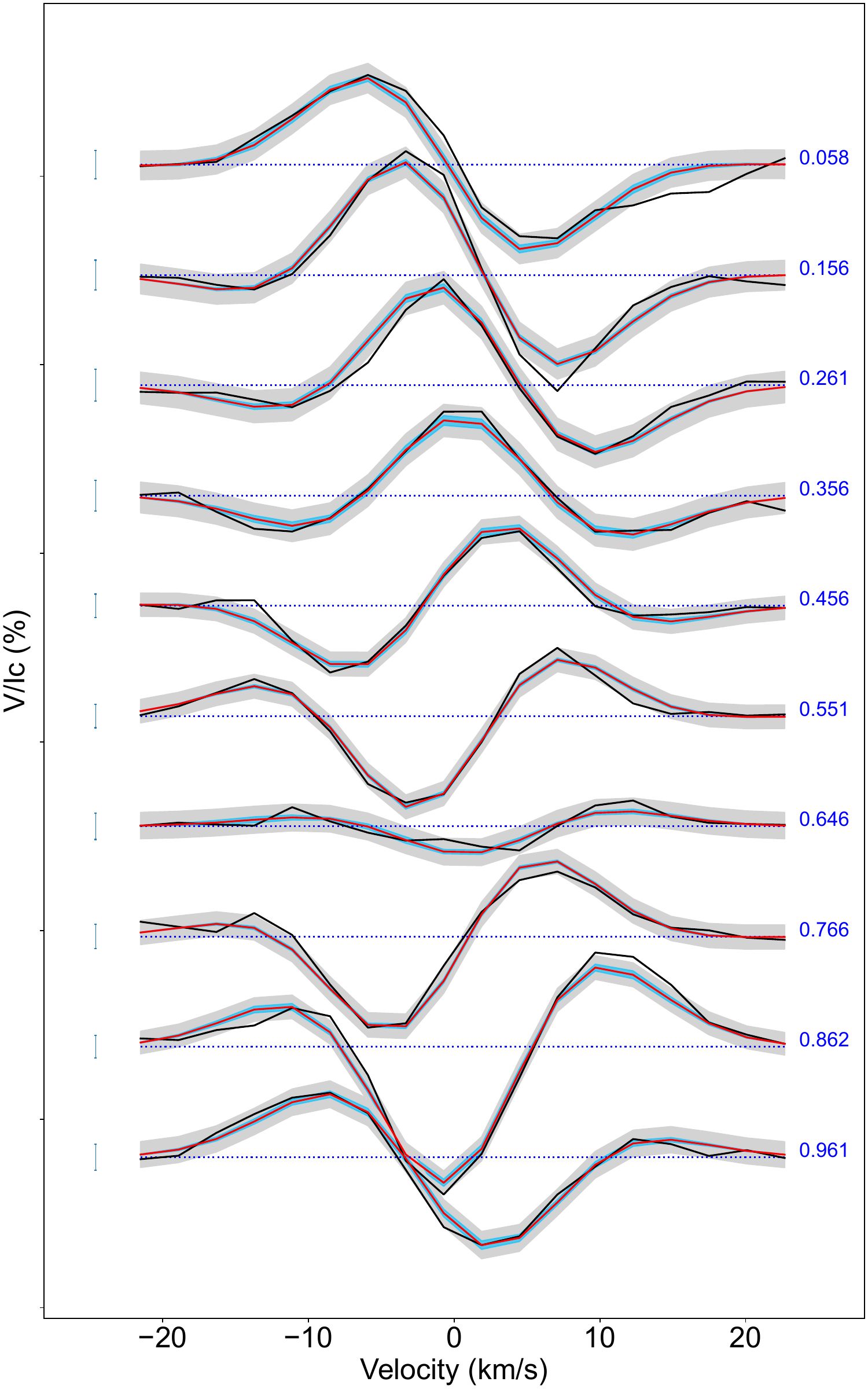}
\caption{\cla{Same as Fig.~\ref{fig:predictive_distribution} but for results obtained using a statistical model with \textbf{a)} a mixture prior consisting of 1000 $\eta$-dependent components $p_c(\mathbf{z})$. and \textbf{b)} a mixture prior consisting of the two components $p_1(\mathbf{z})$ and $p_2(\mathbf{z})$.}}
\label{fig:predictive_distribution_L1L2_comp_subset}
\end{figure*}

\cla{Fig. \ref{fig:predictive_distribution L1L2_comp} and Fig. \ref{fig:predictive_distribution_L1L2_comp_subset} show the corresponding results obtained from statistical models using the two different mixture priors explored in this paper (referred to as Case 2 and Case 3). We observe that the posterior predictive uncertainty is significantly larger using the mixture prior with 1000 $\eta-$dependent components (Case 2)}. 

\FloatBarrier
\section{\cla{Results for spherical harmonic coefficients}}\label{appendix_c}

\begin{figure*}[h!]
\centering
\includegraphics[width=0.49\textwidth]{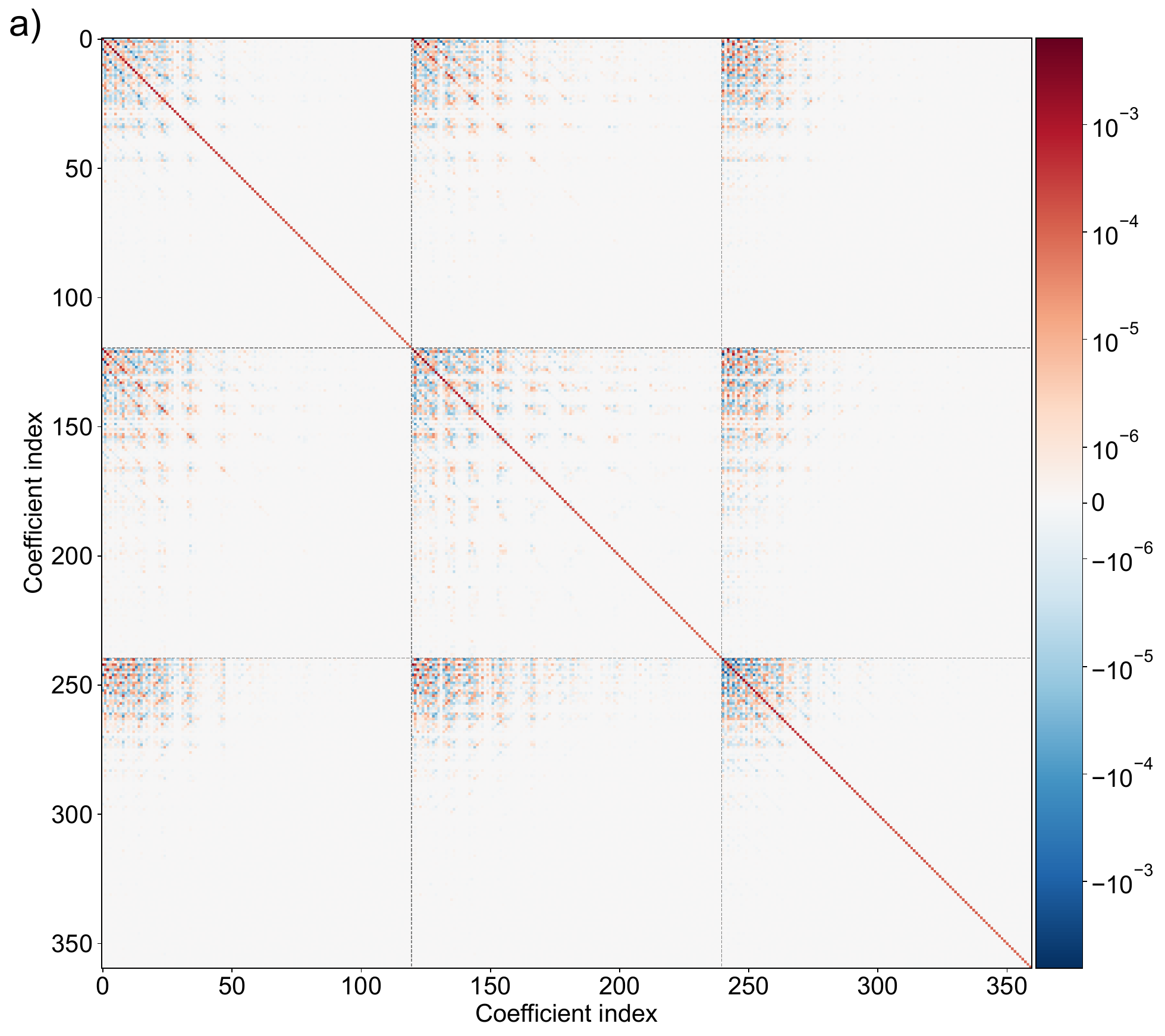}
\includegraphics[width=0.49\textwidth]{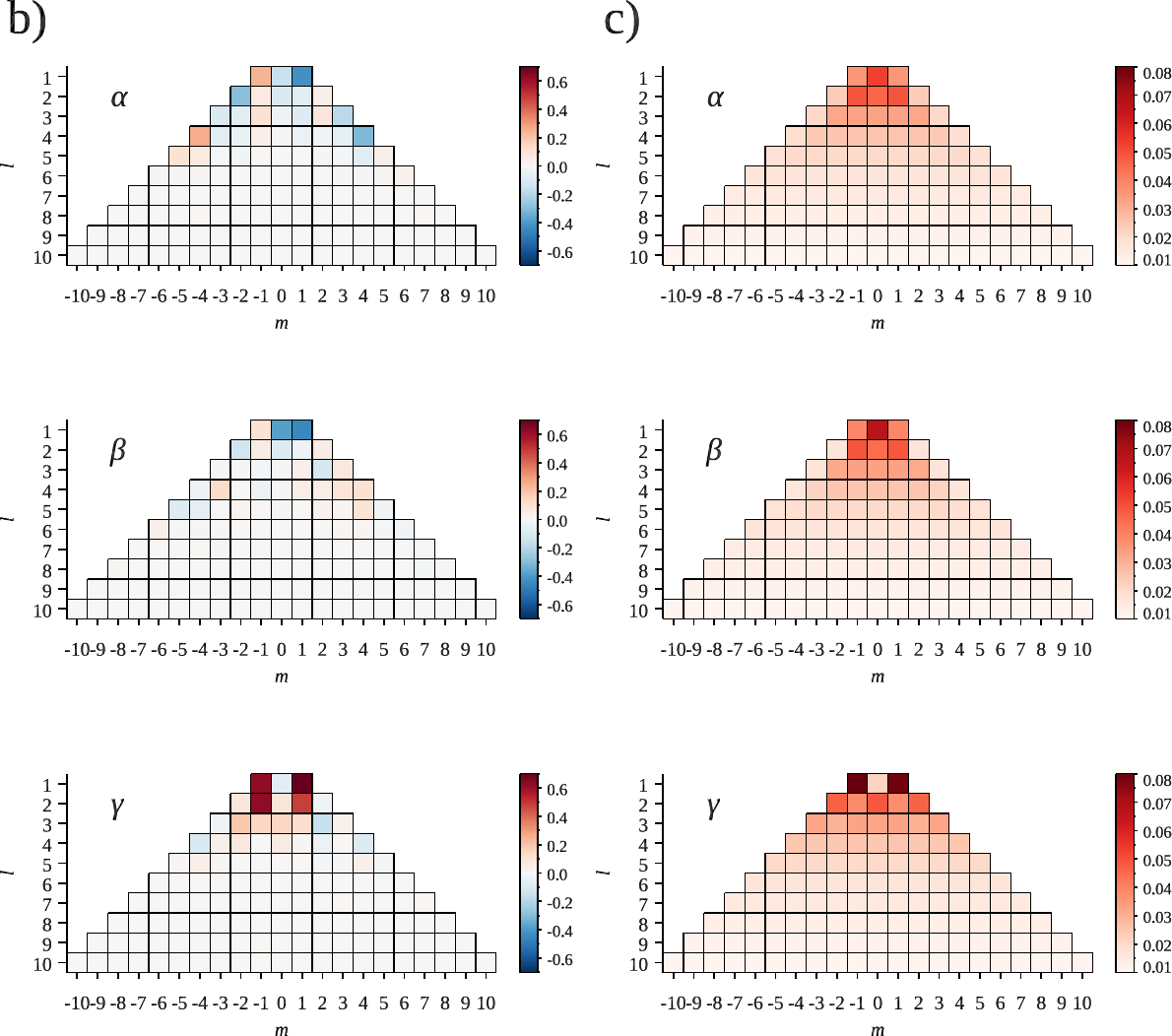}
\caption{\cla{Graphical representation of the spherical harmonic coefficients and their covariances. This plot corresponds to the magnetic field reconstruction results in Fig.~\ref{fig:img_l2}. 
\textbf{a)} Covariance matrix with the three groups of spherical harmonic coefficients, $\alpha$, $\beta$, $\gamma$, stored sequentially in the order of increasing $l$ and $m$ numbers. Dashed lines highlight parts of the covariance matrix corresponding to each group of harmonic coefficients.
\textbf{b)} Mean value of the spherical harmonic coefficients.
\textbf{c)} Standard deviation of the spherical harmonic coefficients.}
}
\label{fig:CoefL2}
\end{figure*}

\begin{figure*}[h!]
\centering
\includegraphics[width=0.49\textwidth]{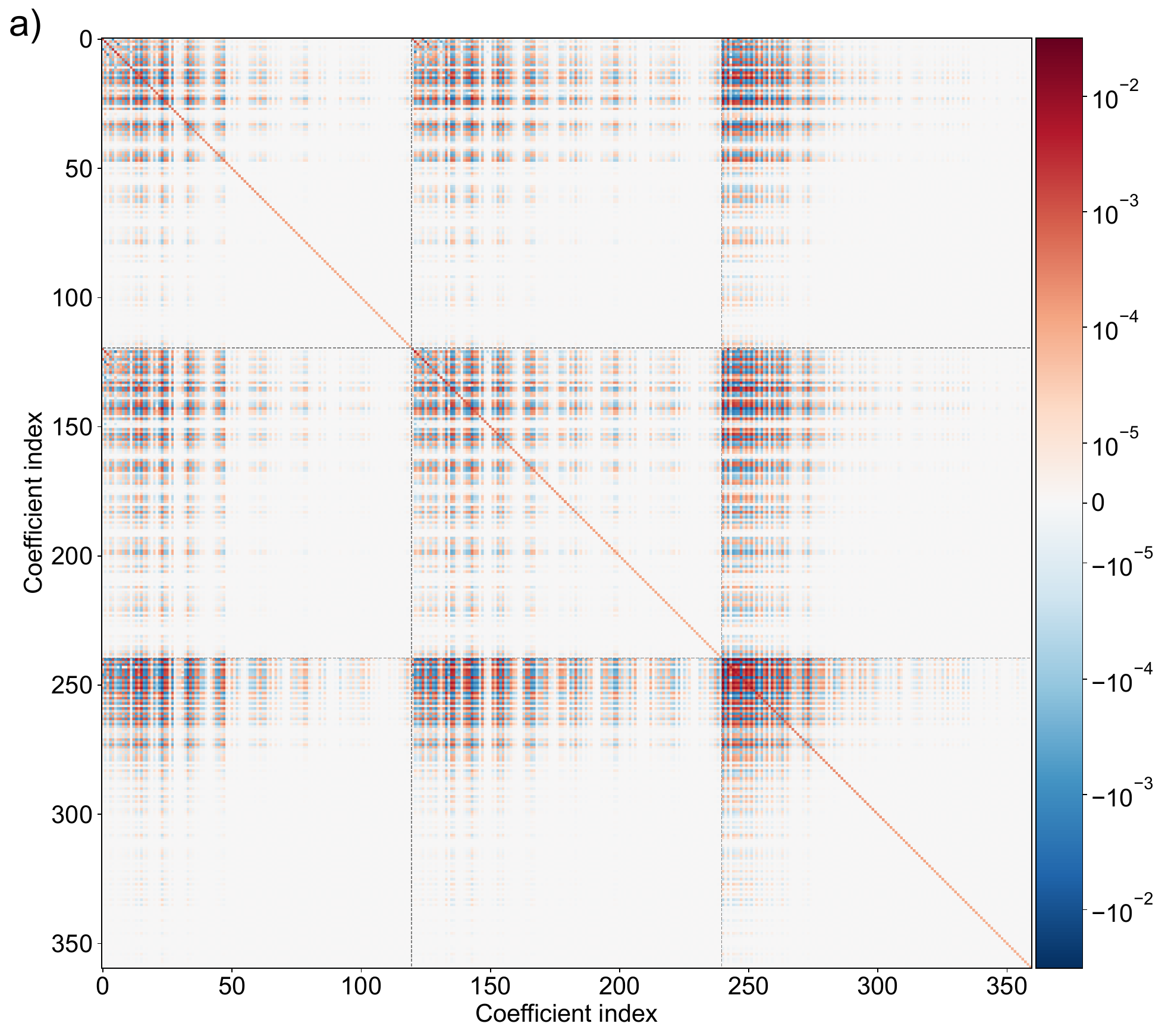}
\includegraphics[width=0.49\textwidth]{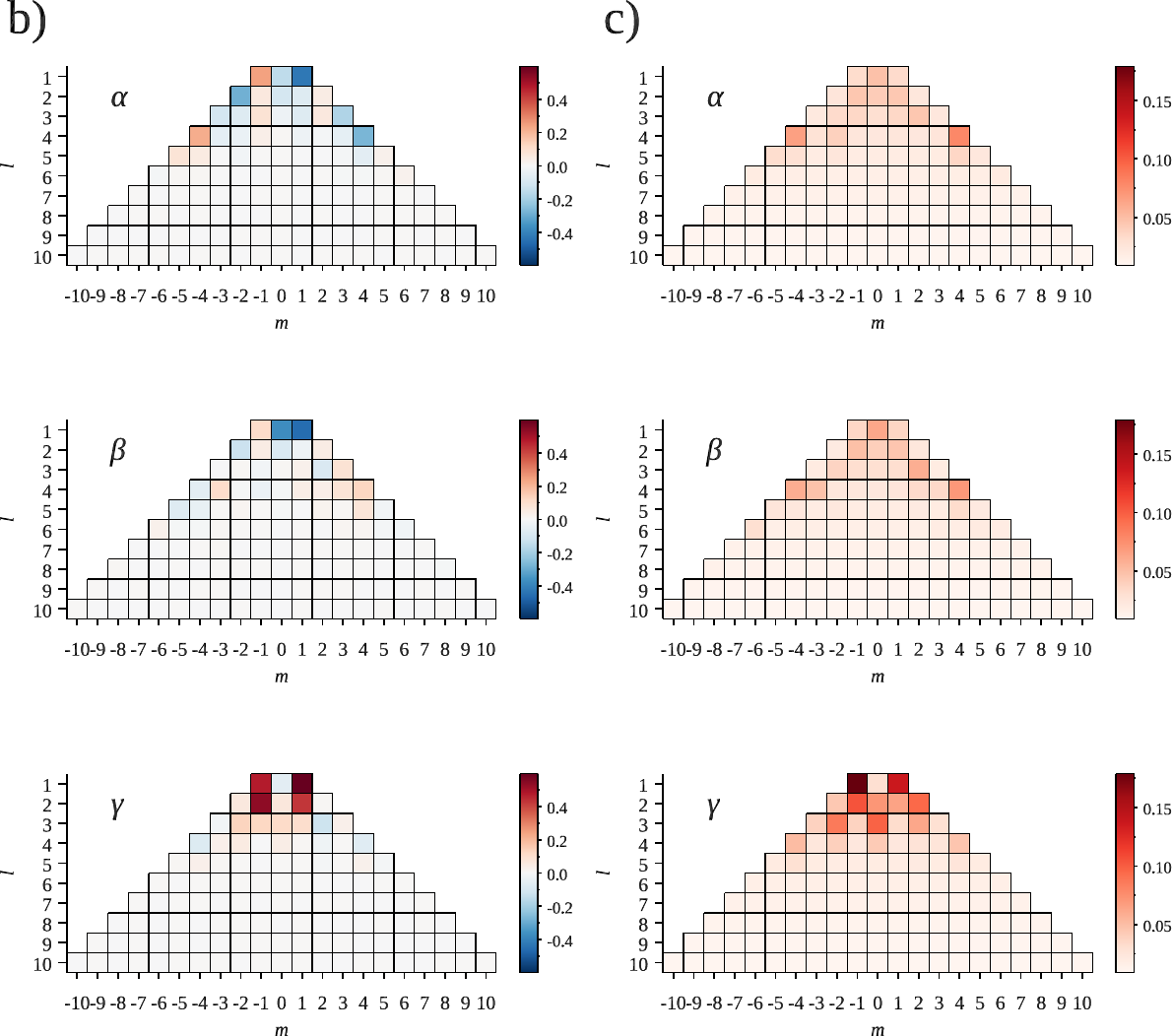}
\caption{\cla{Same as Fig.~\ref{fig:CoefL2} but for the spherical harmonic coefficients corresponding to the magnetic field reconstruction in Fig.~\ref{fig:img_tau}.}}
\label{fig:CoefTau}
\end{figure*}

\begin{figure*}[h!]
\centering
\includegraphics[width=0.49\textwidth]{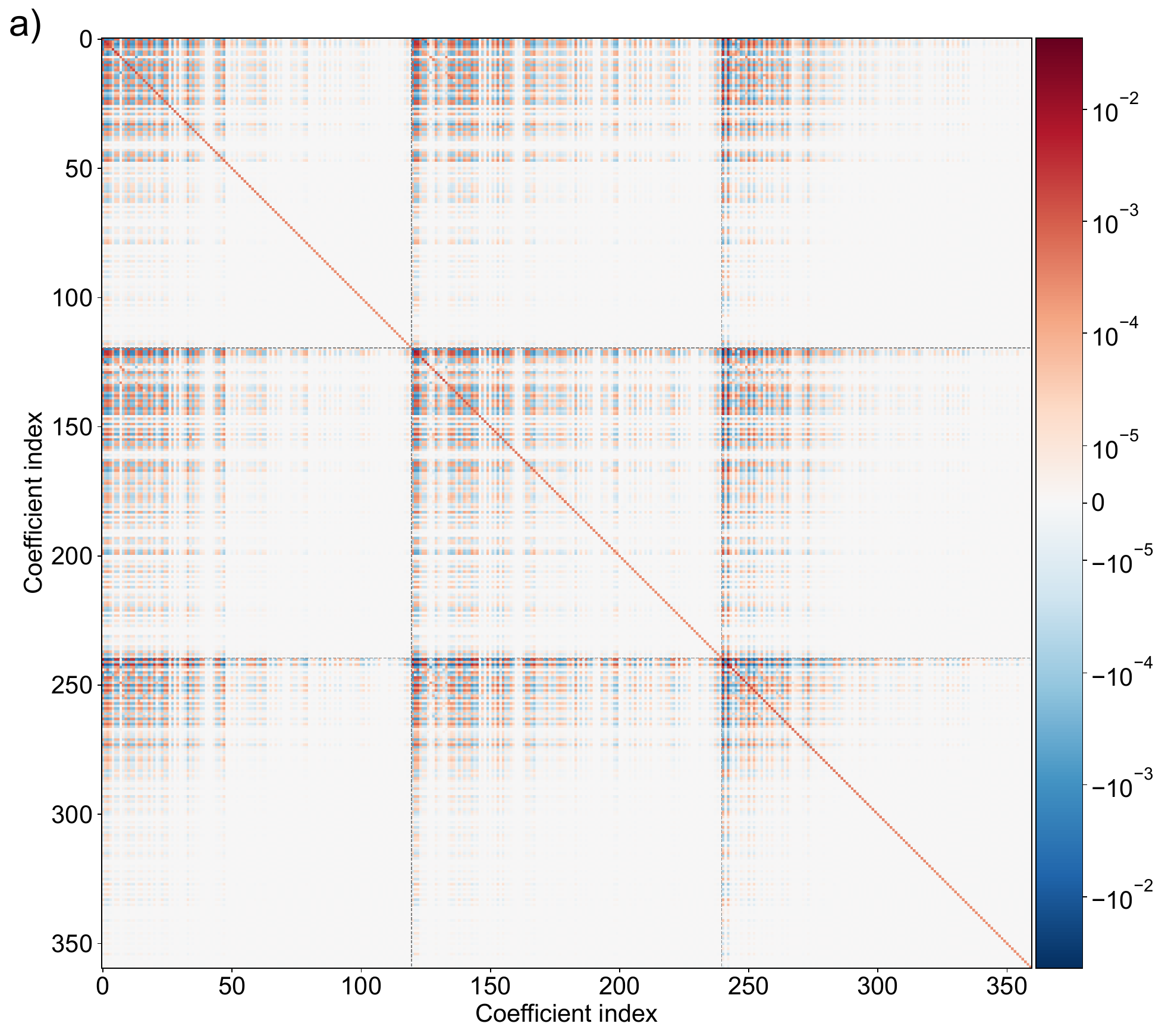}
\includegraphics[width=0.49\textwidth]{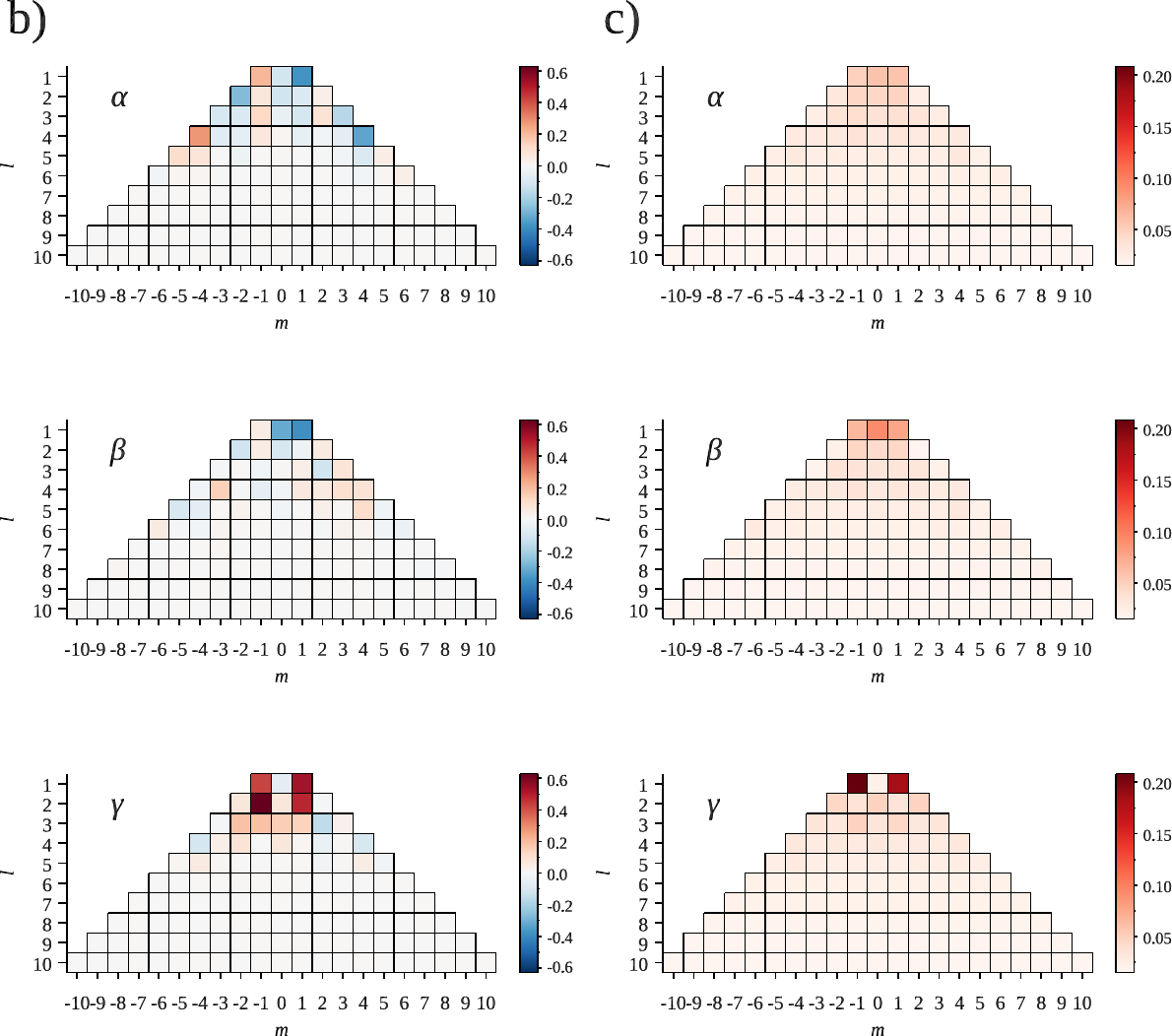}
\caption{\cla{Same as Fig.~\ref{fig:CoefL2} but for the spherical harmonic coefficients corresponding to the magnetic field reconstruction in Fig.~\ref{fig:img_l1l2}.}}
\label{fig:CoefL1L2}
\end{figure*}

\cla{In the main paper, we focus on the posterior magnetic field distribution and the magnetic energy distributions as a function of $l$. The posterior distribution over the spherical harmonic coefficients $\mathbf{z}$ is also readily available but more difficult to interpret. In addition, the values of the coefficients are generally not comparable between studies due to different formulations in terms of e.g. real and imaginary representations of spherical harmonic expansions and different normalisation approaches. For completeness, we present the covariance matrices and triangular plots for the three posterior distributions $p(\mathbf{z}|\mathbf{y})$ presented in this paper. These results can be found in Fig. \ref{fig:CoefL2}, Fig. \ref{fig:CoefTau} and Fig. \ref{fig:CoefL1L2}, corresponding to  Case 1, Case 2 and Case 3, respectively. The uncertainties over the coefficients are dominated by the trend with $l$. The apparent trends with $m$ (e.g. a higher uncertainty for the odd-$m$ toroidal modes for $l=1$ but a lower uncertainty for the same modes for $l=2$) reflect the amplitude of recovered harmonic coefficients and disappear when one considers uncertainties normalised by the absolute value of the coefficients.}

\end{appendix}

\end{document}